\documentclass[12pt,leqno]{article}
\usepackage[utf8]{inputenc}
\usepackage{setspace}
\usepackage{times}
\usepackage{palatino}
\usepackage[left=2.5cm,right=2.5cm,top=2.5cm,bottom=2.5cm]{geometry}
\usepackage{amsmath}
\usepackage{amssymb}
\usepackage{enumerate}
\usepackage[semicolon,round,longnamesfirst]{natbib}
\usepackage[colorlinks=true]{hyperref}
\hypersetup{citecolor=blue}
\usepackage{amsthm}
\usepackage{mathrsfs}
\usepackage{float}
\usepackage{tikz}
\usepackage{amsfonts}
\usepackage{color}
\usepackage{sectsty}
\sectionfont{\centering\textsc}
\subsectionfont{\centering}
\subsubsectionfont{\centering}
\usepackage{subcaption}
\usepackage{appendix}
\usepackage{multirow}
\usepackage{bbm}

\newtheorem{definition}{Definition}
\newtheorem{theorem}{Theorem}
\newtheorem{proposition}{Proposition}
\newtheorem{lemma}{Lemma}

\title{\textsc{\textbf{Grabbing The Forbidden Fruit: \\ Restriction-Sensitive Choice}}\footnote{\scriptsize{This paper previously circulated with the title ``Reactance: a Freedom-Based Theory of Choice''. We are grateful to Miguel Ballester, Roland Bénabou, Francis Bloch, Yves Breitmoser, Franz Dietrich, David Dillenberger, Georgios Gerasimou, Paola Manzini, Marco Mariotti, Pietro Ortoleva, Eduardo Perez-Richet, Evan Piermont and Jean-Marc Tallon for helpful discussions and suggestions, as well as seminar participants at LEMMA Paris 2, Paris 8, BiGSEM, PSE, Sciences-Po, CEC 2021, BRIC 2022, FUR 2022, ESEM 2022 and SAET 2023. N. Boissonnet thanks the support of JICE-InChange which is funded by "Profilbildung 2020", an initiative of the Ministry of Culture and Science of the State of North Rhine-Westphalia. The sole responsibility for the content of this publication lies with the author. A. Ghersengorin thanks the ANR-17-CE26-0003 and the European Research Council (grant 101001694 – IMEDMC and 101040122 - IMD) for their financial support.}}}

\author{Niels \textsc{Boissonnet}\footnote{Bielefeld University. Email: niels.boissonnet@gmail.com.} \hspace{1cm} Alexis \textsc{Ghersengorin}\footnote{University of Bonn. Email: a.ghersen@gmail.com.}}
\date{\normalsize \today} 

\begin{document}

\maketitle
\begin{abstract}
\begin{singlespace}
Restricting individuals' access to some opportunities may steer their desire toward their substitutes, a phenomenon known as the \textit{forbidden fruit effect}. We axiomatize a choice model named \textit{restriction-sensitive choice} (RSC), which rationalizes the forbidden fruit effect and is compatible with the prominent psychological explanations: reactance theory and commodity theory. The model is identifiable from choice data, specifically from the observation of choice reversals caused by the removal of options. We conduct a normative analysis both in terms of the agent's freedom and welfare. We apply our model to shed light on two phenomena: the backfire effect of beliefs and the backlash of integration policies targeted towards minorities.  \\
\end{singlespace}


\end{abstract}

\newpage

``Prohibitions create the desire they were intended to cure.'' 

\hspace*{\fill}  Lawrence Durrell \\

\noindent According to the Italian adage ``\textit{I frutti proibiti sono i più dolci}'' (forbidden fruits are the sweetest), individuals have a natural tendency to place a higher value on things that are prohibited or unavailable. This "forbidden fruit effect" (FFE) \citep{Levesque2018} is frequently emphasized by psychologists as explaining reactions to a wide range of public decisions --- including environmental policy backlash, reluctance to follow nudges, and defiance against government health recommendations --- as well as behaviors arising from shortages caused by natural disasters, strikes, or war.\footnote{\label{fn_expe_evidence}An instance of environmental policy backlash is the ``rolling coal movement'' in the US: in reaction to regulations of car gas emissions, some citizens modified their engine at significant costs to pollute more; see \cite{mazis1973elimination} for the public reaction to a prohibition of a chemical product; see \cite{AradRubinstein2018} for the reaction to nudges; see \cite{Pechman_99} for smoking decisions; see \cite{Hankin1993} for alcohol consumption; see \cite{jansen_07, jansen_08} for eating behaviors; see \cite{bushman_06, Sneegas_98, Varava_15, Gosselt_12} for media choices.} Despite its economic significance, the FFE has received limited attention in economics, and no formal framework has been developed to address it. This paper aims to bridge this gap by studying the implications of the FFE on choice behaviors and proposing a novel choice model that accommodates it.

The FFE poses an intriguing challenge to choice theory as it contradicts the standard model of preference (or utility) maximization, which entails that the agent's choices satisfy the ``Independence of Irrelevant Alternatives'' (IIA) \citep[][property $\alpha$]{Chernoff_54,Sen71}.\footnote{Property $\alpha$ is a weakening of the \textit{Weak Axiom of Revealed Preferences} \citep{Samuelson38}, that is necessary and sufficient to explain a single-valued choice function by the maximization of a linear order \citep[see][]{Sen71}.} To illustrate this point, consider a field experiment studied by \cite{mazis1973elimination} in 1972, where Miami-Dade County banned the use of phosphate detergent. Despite its strong environmental rationale, this decision sparked protests as well as unexpected reactions. In the name of “American freedom”, some consumers with no prior preference for phosphate detergent started buying it in neighboring counties and smuggling it at an additional cost.\footnote{\cite{mazis1973elimination} showed that this astonishing effect on behavior was consistent with the reversal of consumer beliefs: on average, Miami consumers were more likely to praise the efficiency of phosphate detergent than their Tampa county neighbors.} Denoting by $x$ the phosphate detergent in a neighboring county, $y$ the same product in Miami, and $z$ a phosphate-free detergent in Miami, the following choice reversal occurs: $z$ is chosen from the set $\{x,y,z\}$ while $x$ is chosen over $z$ once $y$ is removed, i.e., in the menu\footnote{Note that standard consumer theory cannot generate the FFE. The reason is that the FFE is fundamentally a reaction to the denial of access to options that would not necessarily be chosen when available. It thus results from a change of preferences or from menu-dependent preferences. In contrast, consumer theory posits that an agent maximizes a stable utility over baskets of goods subject to a budget constraint. As a consequence, the consumer's choices satisfy IIA. For instance, the ban of (or a price increase in) an unchosen type of goods cannot modify the consumption choice, contrary to what the FFE predicts. Furthermore, the FFE must not be confused with the existence of a Giffen good, whose consumption increases after a price increase because the income effect dominates the substitution effect.} $\{x,z\}$.

Our goal is to capture the situations in which such choice reversals result from a reaction to the restriction embodied by the denied access to some options (in our example, the phosphate detergent in Miami).\footnote{Note for instance that the \textit{attraction} or the \textit{compromise effects} trigger similar choice patterns. The causes of the choice reversals are, however, different in these cases (see Section \ref{section_lit} for a more detailed comparison).} To that purpose, we characterize a class of choice procedures, named \textit{restriction-sensitive choice} (RSC), that captures the FFE (Section \ref{section:rsc}). According to RSC, the decision-maker (DM) categorizes the options into different \emph{types} (e.g., phosphate and non-phosphate detergents). When deprived of access to sufficiently good options from one type (e.g., phosphate detergent in their local supermarket), the DM may be willing to react by reversing their choice in favor of a less good option of that type (e.g., phosphate detergent in the neighboring county). One common relaxation of IIA and two additional axioms characterize a general version of RSC, where the DM's willingness to react can take any form (Theorem \ref{thm1}). By adding one axiom, we characterize \emph{single-peaked RSC}, where this willingness is single-peaked (Theorem \ref{THMSinglePeaked}), that is, the more the DM is denied access to sufficiently good options of one type, the more she is willing to react, up to a certain point. We show how observed choice reversals can identify the key components of the model (Propositions \ref{RSC_property}–\ref{min_RS_structure}).

We relate our model to the two prominent psychological explanations of the FFE: \textit{reactance theory} \citep{brehm1966theory} and \textit{commodity theory} \citep{Brock1968} (Section \ref{section:psychology}). While according to the former, people's reactions to restrictions or prohibitions result from their attitudes toward freedom of behavior, the latter predicts that the more a commodity is perceived as unavailable or requiring much effort to be obtained, the more it is attractive.  

Section \ref{section:discussion} illustrates how our model suitably captures experimental evidence and discusses possible criticism of our framework. Section \ref{section_lit} compares RSC with the related models from the choice theory literature. Specifically, RSC is a two-step choice model that is a strict subclass of a large class of existing models \citep[see, among others,][]{manzini2007sequentially, masatlioglu2012revealed, cherepanov2013rationalization, horan_2016}. We show that none of these models satisfy RSC's key axioms.

Our model leads to two normative questions: (i) What can be said about the welfare derived by the agent from two distinct options? (ii) How does the agent evaluate the freedom offered by different sets of opportunities? We propose a welfare criterion based on RSC and compare it with existing criteria in the literature (Section \ref{subsec_welfare}), and we axiomatize a measure of freedom on the sets of opportunities for an agent whose choice behavior follows single-peaked RSC (Section \ref{subsec_freedom}).

Finally, we provide two applications of our model (Section \ref{applications}). First, we incorporate RSC in a model that involves the allocation of attention to different media sources, revealing how this can result in a backfire effect on beliefs. Second, we examine the consequences of a government seeking to enforce the integration of a minority group by restricting parents' choice of cultural transmission to their offspring. We demonstrate how these restrictions may lead to a backlash that actually enhances the persistence of a minority.

\section{Restriction-Sensitive Choice}
\label{section:rsc}

\subsection{Framework}

There is a finite set of options $X$, and $\mathcal{X} = 2^X \setminus \emptyset$ denotes the collection of non-empty subsets of $X$. The elements of $\mathcal{X}$ represent the menus of options available to the DM. A \textbf{choice function} $c: \mathcal{X} \longrightarrow X$ associates with each menu the option chosen by the DM in this menu. That is, for any menu\footnote{For simplicity, for a set $\{x_1,\dots,x_k\}$, we write $c\{x_1,\dots,x_k\}$ instead of $c(\{x_1,\dots,x_k\})$.} $A$, $ c(A)\in A$. We assume that menus are exogenously given and do not explicitly model a third party restraining the DM's menu (see Section \ref{section:discussion} for a discussion).


\subsection{Reaction to Restrictions and Comparable Options}

Our analysis is motivated by the influence unchosen options may have on the DM's choices. Specifically, we are interested in cases where the absence of those apparently irrelevant options triggers a \textit{reaction} by the DM. Yet, differentiating such a reaction from standard preference maximization amounts to observing a choice reversal, that is, a violation of the \textit{Independence of Irrelevant Alternatives} axiom \citep[IIA, or property $\alpha$ in][]{Sen71}.\footnote{IIA explicitly requires that removing an ``apparently irrelevant'' option (i.e., not chosen) from a menu must not change the DM's choice: namely, $c(A) = c(B)$ for any $B \subseteq A$ such that $c(A) \in B$.} This motivates the following definition, aimed at capturing when the choice of an option $x$ results from a reaction to the absence of another option $y$.

\begin{definition}\label{def:reaction}
    Let $c$ be a choice function and $x,y\in X$. We say that $x$ \textbf{reacts to the absence of $y$}, relative to $c$, if there exists $z$ such that, $z=c\{x,y,z\}$, and $x=c\{x,z\}$. We denote it\footnote{Our results do not change if the relation $\mathbf{R}^c$ is not defined using a triplet, but any set, i.e.: $x\mathbf{R}^c y$ if there exists a set $A$ such that $x=c(A\setminus\{y\})\neq c(A) \neq y$.} $x \mathbf{R}^c y$.
\end{definition}

Our interpretation of these choice reversals in terms of the FFE is as follows: $x\mathbf{R}^c y$ reveals that the absence of $y$ prompts the DM to choose $x$, even though the option $z$ chosen in the larger menu is available. While these choice reversals allow for competing interpretations, our axiomatization and representation will make ours more suitable.\footnote{In particular, prominently studied phenomena such as the \textit{compromise} or the \textit{attraction} also lead to similar choice patterns, but the interpretation is different (see Section \ref{section_lit} for further details).}

Choosing $x$ as an appropriate reaction to the absence of $y$ should inform the analyst about the DM's perception of comparable features between $x$ and $y$. Returning to our original example, consumers drove to the neighboring county to purchase phosphate-based detergent, not phosphate-free detergent, as the ban only applied to the former. In other words, a DM troubled by the absence of an option may eventually turn to an alternative she deems a suitable potential substitute. Therefore, understanding the subjective perception of which options are comparable in terms of reaction behavior is essential for a revealed preference analysis. We interpret $x \mathbf{R}^c y$ as revealing that the DM perceives $x$ and $y$ as being \textit{reaction-comparable} (in short, \textit{comparable}). We capture the different classes of comparable options by extending this notion of comparability as follows.

\begin{definition}\label{def:similar}
    Let $c$ be a choice function and $x,y\in X$. We say that $x$ and $y$ are \textbf{\emph{reaction-comparable}} (in short, \textbf{\emph{comparable})}, denoted $x\approx y$, if $x=y$ or there exists $x_1,\dots, x_{n}\in X$ such that 
    \begin{align*}
    (x\mathbf{R}^cx_1\vee x_1\mathbf{R}^cx) \wedge (x_1\mathbf{R}^cx_2\vee x_2\mathbf{R}^cx_1)\wedge\dots \wedge (x_n\mathbf{R}^cy\vee y\mathbf{R}^cx_n).
    \end{align*}
    The relation $\approx$ is an equivalence relation (reflexive, transitive, and symmetric), for which $X_\approx$ is the set of its equivalence classes. We refer to elements of $X_\approx$ as \emph{comparability classes}. 
\end{definition}

The comparability relation $\approx$ is the reflexive and transitive closure of the symmetric relation that links $x$ and $y$ whenever $x\mathbf{R}^c y$ or $y\mathbf{R}^cx$. A class in $X_\approx$ contains options that are all linked to each other (possibly indirectly) through reaction behaviors. As the next section clarifies, the comparability class of an option $x$ is composed solely of the options that the DM may choose as a substitute to \emph{react to the absence of} $x$. One possible interpretation of $\approx$ is \emph{similarity}, under the disputable requirement that this similarity relation be transitive.

\subsection{Representations}
\label{section:representations}

In our model, restrictions are captured by reduced sets of feasible options. Accordingly, expanding a menu should weaken the motive to react. Specifically, suppose $x$ is chosen from menu $A$. Expanding $A$ by adding a set of options $B$ in which $x$ is also chosen should not induce any change in the choice. Otherwise, the reversal from $A\cup B$ to $A$ would be driven by the loss of an option from $B\setminus A$, which should prevent $x$ from being chosen in $B$. This consistency is precisely what our first axiom, a standard relaxation of the \textit{Weak Axiom of Revealed Preference}, requires.\footnote{It was already present in \cite{Sen71}, named property $\gamma$, and was later popularized by \cite{manzini2007sequentially} under the name \textit{Expansion}.}

\paragraph{\textsc{Expansion} (Exp).\label{Exp}}
\textit{For any $x\in X$, $A,B\in \mathcal{X}$, if $x = c(A) = c(B)$, then $x = c(A\cup B)$.} \\

By definition of $\mathbf{R}^c$, there cannot be any intransitive binary choices among options that are all in different comparability classes. Indeed, if $x=c\{x,y\}$, $y=c\{y,z\}$, and $z=c\{x,z\}$, then depending on $c\{x,y,z\}$, we would conclude that $x\mathbf{R}^cz$, $z\mathbf{R}^cy$, or $y\mathbf{R}^cx$, implying that at least two options are in the same comparability class. The second axiom requires that this also holds for options that are all comparable to each other. In particular, combined with \textbf{Exp}, it implies that if $x$ reacts to the absence of $y$, this should be revealed by a third option $z$ (such that $z=c\{x,y,z\}$) that is \emph{not} comparable to $x$ and $y$.\footnote{To see why, note that if $c$ satisfies \textbf{Exp}, then whenever $x\mathbf{R}^c y$, there is a cycle of binary choices with a third option $z$: $x=c\{x,z\}$, $y=c\{x,y\}$ and $z=c\{y,z\}$. Otherwise, this would mean that one option is chosen over the two others, and \textbf{Exp} would imply that this option is also chosen in the bigger set with the three options.} When caused by the FFE, intransitive binary choices result from the DM re-evaluating her priorities: she turns away from $z$ because the absence of $y$ drives her to choose an option comparable to $y$. Yet, if $z$ were also comparable to $y$, it would align with this new priority, and the DM should not turn away from it. More concretely, a ban on phosphate detergent in Orlando should not lead Miami residents to drive to Tampa to obtain the same product if it is still available in Miami.

\paragraph{\textsc{No-Reaction Comparability} (NRC).} For all $x,y,z$ such that $x\approx y\approx z$, if $x=c\{x,y\}$ and $y=c\{y,z\}$, then  $x=c\{x,z\}$. \\

This axiom, coupled with \textbf{Exp}, is equivalent to satisfying IIA within each comparability class, and thus to the existence of a well-ordered criterion (or preference) that rationalizes the DM's choices within these classes. For instance, a comparability class may contain several instances of the same product with different satisfaction levels due to differential effort, cognitive, or monetary costs. Therefore, while our deliberate agnosticism regarding the subjective reaction comparability among options prevents us from explicitly modeling this criterion, \textbf{Exp} and \textbf{NRC} allow us to uncover it from choices.

The FFE stems from the idea that the choice of an option $x$ is guided both by its intrinsic desirability and the psychologically motivated response to the absence of comparable options. The last axiom requires that the expression and magnitude of these two forces depend only on the availability of comparable options and are independent of the feasibility of non-comparable options. Specifically, if two comparable options $x$ and $y$ are such that for a non-comparable option $z$, we observe that $x$ is chosen over $z$ ($x=c\{x,z\}$) while $z$ is chosen over $y$ ($z=c\{y,z\}$), then it must be that, whenever $y$ is chosen over a non-comparable option $t$ ($y=c\{y,t\}$), so is $x$ ($x=c\{x,t\}$). The interpretation is as follows: if the forces compelling the DM to choose $x$ over $z$ were sufficiently strong to overcome her preference for $z$ over $y$, then these forces should consistently lead the DM to choose $x$ over any non-comparable option $t$ that is not even chosen over $y$. 

\paragraph{\textsc{Independent Reaction} (IR).} For all $x,y,z,t$ such that $t\not\approx x\approx y\not\approx z$: if $x=c\{x,z\},$ $z=c\{y,z\}$, and $y=c\{y,t\}$, then $x=c\{x,t\}$.
\\

Consider an illustrative example. Suppose that phosphate detergent is banned in Miami (option $y$) but not in Tampa (option $x$), and a Miami resident chooses $x$ rather than continue buying the high quality phosphate-free alternative she used to consume $z$ (formally, $x = c\{x,z\}$ and $z = c\{x,y,z\}$). According to our axiom, if this consumer prefers phosphate detergent in Miami $y$ over another lower-quality phosphate-free alternative $t$ ($y = c\{y,t\}$), she would maintain this preference when the phosphate detergent is banned in Miami, opting to drive to Tampa for $x$ rather than switching for $t$ ($x = c\{x,t\}$).

\paragraph{Restriction-Sensitive Choice.} 
We later add a final axiom that imposes additional structure on the choice behaviors and yields a representation that, we argue, suitably captures the FFE (see Section \ref{section:psychology} for our comparison with the psychology literature). However, the three previous axioms are already necessary and sufficient to obtain a representation. We now present this first result for both its independent interest and its pedagogical value. Anticipating our main representation, we name this more general representation \emph{restriction-sensitive choice}.

\begin{definition}\label{RSC}
    A choice function $c$ is a \textbf{\emph{restriction-sensitive choice}} (RSC) if there exist a partition $\mathcal{T}$ of the options into \emph{types} and two linear orders\footnote{A \emph{linear order} defined on a set $E$ is a complete, transitive and antisymmetric binary relation.} on $X$ --- a \emph{welfare preference} $\succsim_1$, and a \emph{reaction preference} $\succsim_2$ --- such that, for any menu\footnote{The notation $\max(B, \succsim)$ refers to the maximal element in $B$ according to the linear order $\succsim$, i.e., $\{x \mid x \succsim y, \forall y \in B\}$. Accordingly, $\min(B,\succsim)$ refers to the minimal elements. With a linear order, these sets are singletons; thus, we slightly abuse notations and interchangeably use these notations to refer to the set or the unique element in that set.} $A$,
    \begin{align}
        & c(A) =\max( d(A),\succsim_2), \label{RSC_choice} \\
        \text{where} \qquad & d(A) = \bigcup\limits_{T \in \mathcal{T}}\max( T \cap A,\succsim_1).\label{RSC_choice2}
    \end{align}
    In this case, we say that $\langle \mathcal{T}, \succsim_1, \succsim_2 \rangle$ is an \emph{\textbf{RS-structure}} that \emph{rationalizes} $c$.
\end{definition}

\begin{theorem}
    A choice function $c$ satisfies \textbf{Exp}, \textbf{NRC}, and \textbf{IR} if and only if it is an RSC.
    \label{thm1}
\end{theorem}
According to RSC, the DM categorizes the options into different \emph{types} which form a partition, and by (\ref{RSC_choice}) the choice results from a two-step maximization.\footnote{Section \ref{section_lit} discusses in detail the relationship between RSC and the existing literature on multiple-step maximization choice models.} The DM first retains the top available options from each type according to the \emph{welfare preference} $\succsim_1$, forming the set $d(A)$. Then, the options within $d(A)$ are evaluated through the \emph{reaction preference} $\succsim_2$.

Our interpretation of RSC in light of the FFE is as follows. Types gather options that the DM deems comparable (see Proposition \ref{proposition:types} for the connection with $X_\approx$). When choosing among options from the same type, the DM only uses the welfare preference $\succsim_1$, which represents the DM's intrinsic satisfaction, or material welfare. When choosing among options from different types, the DM uses the reaction preference $\succsim_2$, which, in addition to welfare considerations, also includes the DM's propensity to react to restrictions. In the next section, we detail how this interpretation relates to the main psychological theories explaining the FFE: reactance theory and commodity theory (Section \ref{section:psychology}). 

A final axiom will add structure to RSC and strengthen this interpretation in terms of the FFE. At this stage, however, other plausible interpretations of RSC exist. A first interpretation views each set $T \in \mathcal{T}$ as a category of alternatives that are easy to compare -- e.g., $T$ contains different iPhone models and $T'$ different Galaxy models. The agent uses the first preference relation $\succsim_1$ for within-category comparisons and the second preference relation $\succsim_2$ for comparisons across categories. Across-category comparisons may be more difficult, and thus $\succsim_2$ may differ from $\succsim_1$.\footnote{We thank one referee for proposing this alternative interpretation.} Another interpretation would be that the DM explicitly uses different criteria to compare within and across categories. For instance, each $T \in \mathcal{T}$ may include every potential candidate for an election from the same party. The DM first ranks candidates according to $\succsim_1$ by their likelihood of winning their party's primary if they run for election. Then, the DM orders candidates across parties by her intrinsic preference $\succsim_2$.

Before introducing the last axiom, we first explain how RSC can generate reactions to restrictions (i.e., $x\mathbf{R}^cy$). Observing $x\mathbf{R}^cy$ not only entails that the reaction preference misaligns with the welfare one ($\succsim_2 \neq \succsim_1$), it requires a specific structure of this misalignment.\footnote{Note that if $\succsim_2 = \succsim_1$, the DM's choices result from maximizing a single preference.} Given that $\succsim_1$ first selects options within types but never serves as a criterion to choose options across types afterward, $\succsim_2$ must be misaligned with $\succsim_1$ on options of the same type. Suppose now that $x$ and $y$ belong to the same type $T$, $y\succ_1 x$, but $x\succ_2 y$. If for any $z \notin T$, whenever $x\succ_2 z$, it is also the case that $y\succ_2 z$, then for any menu where $y$ is not chosen, $x$ would not be chosen either if it were available instead of $y$. Hence, this misalignment does not trigger any reaction to restriction. This leads to the next proposition, which characterizes the conditions under which an RS-structure gives rise to choice reversals, and thus reactions to restrictions. It thereby corroborates our earlier interpretation of $x\mathbf{R}^c y$, that is, observing a reaction to a restriction means that the absence of an unchosen option leads the DM to reverse her choice in favor of a lower-ranked option (according to the welfare preference) that is from the same type.

\begin{proposition}\label{RSC_property}
    Suppose $c$ is rationalized by an RS-structure $\langle \mathcal{T}, \succsim_1, \succsim_2 \rangle$. For any $x,y \in X$, $x\mathbf{R}^c y$ if and only if there exists $T \in \mathcal{T}$ and $z\notin T$, such that $x,y \in T$, $y\succ_1x$ and $x \succ_2z\succ_2 y$.
\end{proposition}

\paragraph{Single-Peaked RSC.} While RSC leaves open the possibility that reactions to restrictions occur, it does not put any structure on when and how they could happen. We would like to specify further the structure of the misalignment in a way that seems natural to us and that conforms with the psychology literature, as we explain next (Section \ref{section:psychology}). Specifically, one could expect the DM's propensity to react to be in proportion to the severity of the restriction. That is, as restrictions intensify and the best available option of a given type worsens (according to $\succsim_1$), the DM becomes more willing to react. This increasing propensity to react may persist only up to a certain point, beyond which welfare costs would exert a counterbalancing effect. 

One additional axiom allows us to capture this monotonic structure of the reaction. Consider three options $x,y,z$ from the same comparability class such that $x$ is chosen over $y$ and $y$ over $z$, that is, $x=c\{x,y\}$ and $y=c\{y,z\}$. Suppose that $x$ reacts to the absence of some option $t$ (i.e., $x\mathbf{R}^c t$), so that the DM has reaction-related motives when she chooses $x$. The question then is how strong those motives are when she chooses $y$ or $z$. Our axiom requires that they must be at least as strong whenever the choice of $z$ reacts to the absence of $y$. Formally, if $z\mathbf{R}^c y$, then it must be that whenever $x$ is chosen over a non-comparable option $u$, so is $y$. The explanation is as follows: $z\mathbf{R}^c y$ reveals that the DM is willing to react even more strongly when the middle option $y$ is no longer available by choosing the bottom option $z$. The axiom then requires that the DM's propensity to react when choosing $y$ must not be weaker than when choosing $x$.

\paragraph{\textsc{Single-Peaked Reaction} (SPR).} For any $x,y,z\in X$ such that, $x\mathbf{R}^ct$ for some $t\in X$, $x \approx y \approx z$, and $x=c\{x,y\}$, $y=c\{y,z\}$: if $z\mathbf{R}^cy$, then for all $u\not\approx x$, $x=c\{x,u\}\implies y=c\{y,u\}$. 
\\

SPR, together with the previous axioms, lead to what we call \emph{single-peaked RSC}. To state our result, we first introduce the notion of single-peakedness of one linear order with respect to another \citep[c.f.][]{Black1948, Arrow1963}. For any linear order $\succsim \subset E^2$, $F\subset E$, and $a,b \in F$, we denote as $[a,b]^\succsim_F = \{x\in F \colon a \precsim x \precsim b\}$ the ``interval'' of elements in $F$ bounded by $a$ and $b$.

\begin{definition}
    Let $\succsim_1$, $\succsim_2 \subset E^2$ be two linear orders and $F\subset E$. We say that $\succsim_2$ is \textbf{single-peaked with respect to $\succsim_1$ on the interval} $[a,b]^{\succsim_1}_F$ if for any $x,y,z\in [a,b]^{\succsim_1}_F$: $x\prec_1 y \prec_1 z$ implies that $y\succ_2 \min (\{x,z\}, \succsim_2)$.
\end{definition}

\begin{definition}\label{RSC}
    An RSC $c$ is \textbf{\emph{single-peaked}} if there exists a rationalizing RS-structure $\langle \mathcal{T}, \succsim_1, \succsim_2 \rangle$ that satisfies the following condition: for any $T \in \mathcal{T}$, there exists $x_T^\star \in T$ such that,
    \begin{enumerate}[(i)]
        \item $\succsim_2 = \succsim_1$ on $\left[x_T^\star, \max(T,\succsim_1)\right]^{\succsim_1}_T$, 
        \item $\succsim_2$ is single-peaked with respect to $\succsim_1$ on $\left[\min(T,\succsim_1), x_T^\star\right]^{\succsim_1}_T$.
    \end{enumerate}
    In this case, we say that $\langle \mathcal{T}, \succsim_1, \succsim_2 \rangle$ is a \emph{single-peaked RS-structure}.
\end{definition}

\begin{theorem}\label{THMSinglePeaked}
    An RSC $c$ satisfies \textbf{SPR} if and only if it is single-peaked. \label{THMSinglePeaked}
\end{theorem}

According to a single-peaked RSC, $x^\star_T$ is the threshold below which restrictions may trigger reactions, as $\succsim_2 = \succsim_1$ above it. In other words, the DM feels \textit{legitimate} to have access to at least one option from each type above the type's threshold. Once the DM starts reacting to restrictions because only options below $x^\star_T$ are available, then the more she is denied access to good options of type $T$, the stronger is her propensity to react and choose an option of this type. This, however, holds up to a certain point, the peak of $\succsim_2$ in the interval $\left[\min(T,\succsim_1), x_T^\star\right]^{\succsim_1}_T$. Beyond this peak, restriction-related motives are still relevant, but they tend to decrease as intrinsic satisfaction decreases with the restriction.

\paragraph{Menu dependence.} RSC's two-step maximization procedure (\ref{RSC_choice}) and  (\ref{RSC_choice2}) typically generates menu-dependent preferences. We can reformulate the representation as the maximization of a single function that directly depends on the menu to make this dependence more explicit. Given a choice function $c$ rationalized by a single-peaked RS-structure $\langle \mathcal{T}, \succsim_1, \succsim_2 \rangle$, there exist a \textit{utility index} $u:X\mapsto \mathbf{R}$, a \textit{threshold function} $\beta:\mathcal{T}\mapsto \mathbb{R}$, and a \textit{reaction index} $\mathcal{R}: \mathcal{T} \times \mathbb{R} \to \mathbb{R}_+$ such that 
\begin{align}
    c(A)=\arg\max_{x\in A}\Bigg[u(x)+\mathcal{R}_{\mathcal{T}(x)}\Big(\max u(\mathcal{T}(x)\cap A)\Big)\cdot\mathbbm{1}_{\{\beta(\mathcal{T}(x))> \max u(\mathcal{T}(x)\cap A\}}\Bigg], \label{eq:menu_dep}
\end{align}
where $\mathcal{T}(x)$ is the type of option $x$, and $\mathcal R_T(\cdot) = \mathcal R(T, \cdot)$. The utility index $u:X\mapsto \mathbf{R}$ captures the genuine utility of each option (it represents $\succsim_1$). The threshold encodes the minimal utility the DM feels entitled to enjoy for each type (it maps $T$ to $u(x_T^\star)$). Finally, the reaction index quantifies the additional value the DM attributes to an option $x$ when she is willing to react to the absence of sufficiently good options from its type $\mathcal{T}(x)$ (i.e., options whose utilities are higher than $\beta(\mathcal{T}(x))$). The mapping $r \mapsto r + \mathcal{R}_T( r)\mathbbm{1}_{\{\beta(T) > r\}}$, defined on the image $u(X)$, represents $\succsim_2$, and thus is quasi-concave. The representation \eqref{eq:menu_dep} makes explicit that the value of option $x$ depends directly on the best option of the same type of $x$ that is available in $A$.

\subsection{Psychological Interpretations}
\label{section:psychology}

Our model is in line with the two prominent explanations of the forbidden fruit effect in psychology: \textit{reactance theory} \citep{brehm1966theory} and \textit{commodity theory} \citep{Brock1968}.\footnote{\label{fn_theories}See \cite{rosenberg_siegel18} for a review of psychological reactance theory, and \cite{Lynn_91} for a review of commodity theory.}

\paragraph{Reactance: A freedom-based theory of choice.} Psychological reactance theory posits that individuals may perceive certain restrictions as a threat to their freedom and consequently choose a dominated option as a way to \textit{restore} the eliminated freedom. In our model, each type embodies a specific freedom.\footnote{Psychologists emphasize that reactance reflects an attempt to restore the loss of concrete freedoms, that is, freedoms to choose diverse types of options. ``Contrary to some interpretations (e.g., \cite{dowd1975distributive}), the freedoms addressed by the theory are not 'abstract considerations', but concrete behavioral realities.'' \citep[p.12]{brehm2013psychological}} The welfare preference $\succsim_1$ represents the intrinsic satisfaction, or material welfare, of the agent. For instance, buying phosphate laundry in the supermarket next door is less costly than getting it in a supermarket in a neighboring county, but both options may be perceived as materializing a similar freedom. As the intrinsic satisfaction of the best option available within a type $T$ diminishes, the freedom embodied by $T$ is reduced. The threshold $x^\star_T$ marks the frontier between the restrictions that the DM considers legitimate to have and those that threaten the freedom embodied by $T$. As long as options with a satisfaction level at least as good as $x^\star_T$ are available, no freedom concern is activated. Hence, the reaction preference $\succsim_2$ aligns with $\succsim_1$ for these options. Conversely, if the welfare of the best option available in $T$ is below $x^\star_T$, the DM deems that they do not have access to a sufficiently good option regarding the freedom associated with $T$. They may then be prone to react, which increases their willingness to choose an option from $T$, although this option provides a lower intrinsic satisfaction. This is how they ``restore'' the eliminated freedom, which is captured by the misalignment between $\succsim_1$ and $\succsim_2$. The requirement that $\succsim_2$ be single-peaked with respect to $\succsim_1$ reflects that the DM is increasingly more willing to react as a particular freedom is further restricted. However, this increasing reaction holds up to a certain point beyond which the welfare cost is too significant to be always offset by freedom concerns \citep[see][for evidence of the single-peakedness of reactance-driven reactions]{rosenberg_siegel18}. 

\paragraph{Commodity Theory: When more constraint increases the value of options.} Commodity theory claims that the DM relates the value of a commodity to the perceived effort required to obtain it \citep{Brock1968}. Under this interpretation, types gather options that the DM considers similar but with different satisfaction levels, or material welfare (as captured by $\succsim_1$). For instance, a type may gather several instances of the same product (e.g., phosphate laundry) with different associated costs (e.g., traveling or effort costs). The threshold $x^\star_T$ captures the level of satisfaction above which further restrictions on the type $T$ are not perceived as requiring unreasonable additional effort to obtain the commodity. Alternatively, one can interpret $x^\star_T$ as capturing the level of satisfaction below which the additional effort to get options of type $T$ becomes \textit{salient}, thus attracting the DM's attention. In that sense, our model is closely related to the literature on consideration sets (see Section \ref{section_lit} for a more detailed comparison with the existing literature). When for a type $T$ no option above the threshold is available, the best alternative option available from $T$ becomes a ``forbidden fruit'' \citep[e.g.,][]{Levesque2018}, and thus all the more attractive. This is captured by $\succsim_2$, which adds an intrinsic restriction-related pleasure on top of the welfare (see the function $\mathcal{R}$ in the representation \eqref{eq:menu_dep}). The single-peakedness of $\succsim_2$ is interpreted along the same line as for reactance: desire may increase with the degree of the effort, but possibly only up to a certain point.

\subsection{Identification}

How can we identify a rationalizing (single-peaked) RS-structure? First, consistently with our interpretation of the similarity relation, the equivalence classes $X_\approx$ can be used to determine the types. Furthermore, it also constrains the types that can be constructed. 

\begin{proposition}\label{proposition:types}
    If $c$ is a (single-peaked) RSC, then there exists a (resp.\ single-peaked) RS-structure $\langle \mathcal{T}, \succsim_1, \succsim_2 \rangle$ that rationalizes $c$ and such that $\mathcal{T}=X_\approx$. Furthermore, for any rationalizing RS-structure $\langle \mathcal{T}', \succsim'_1, \succsim'_2 \rangle$, if $x \approx y$, then there exists $T\in \mathcal{T}'$ such that $x,y\in T$.
\end{proposition}

Second, the key features of single-peaked RSCs can also be easily deduced from the relation $\mathbf{R}^c$. We can identify, for each type $T$, the threshold $x^\star_T$ as well as the peak of $\succsim_2$ on $[\min(T,\succsim_1), x^\star_T]^{\succsim_1}_T$, that is, $\max\big([\min(T,\succsim_1), x_T^\star]^{\succsim_1}_T, \succsim_2\big)$.

\begin{proposition}\label{min_RS_structure}
     If $c$ is a single-peaked RSC, then there exists a rationalizing single-peaked RS-structure $\langle \mathcal{T}, \succsim_1, \succsim_2 \rangle$ such that $\mathcal{T} = X_\approx$, and for any $T\in \mathcal{T}$,
     \begin{enumerate}
         \item $\displaystyle x^\star_T=\min\big(\{x\in T\mid \nexists y, x\mathbf{R}^cy\}, \succsim_1\big)$,
         \item $\displaystyle\max\big([\min(T,\succsim_1), x_T^\star]^{\succsim_1}_T, \succsim_2\big)= \max \big(\{x \in T \mid \nexists y,  y\mathbf{R}^c x\},\succsim_1\big).$
     \end{enumerate}
     We define such a single-peaked RS-structure as \textbf{\emph{minimal}}.
\end{proposition}

According to Propositions \ref{proposition:types} and \ref{min_RS_structure}, one can identify a minimal single-peaked RS-structure with the relation $\mathbf{R}^c$. Indeed, note that once types are determined, then the relevant parts of the linear orders $\succsim_1$ and $\succsim_2$ can be deduced easily: $\succsim_1$ from binary choices within types, and $\succsim_2$ from binary choices across types. Furthermore, minimal single-peaked RS-structures have the following properties that corroborate our interpretation: (i) in a type $T$, any option above $x^\star_T$ is never observed to react to the absence of any other option; (ii) any option below $x^\star_T$ reacts to the absence of some options (Lemma \ref{Lemmarelou1} in Appendix \ref{app:proof_SPR} actually shows that every option below $x^\star_T$ reacts to the absence of $x^\star_T$ itself); and (iii) the removal of options below $\displaystyle\max\big([\min(T,\succ_1), x_T^\star]^{\succ_1}_T, \succ_2\big)$ never triggers any reaction, that is, it is the option that triggers the most restriction-related reaction.

\section{Discussion and Related Literature}

\subsection{Discussion of the Model}
\label{section:discussion}

\paragraph{Relation to experimental works.}

Our approach is consistent with the large body of experimental work documenting the FFE using \textit{between-subject} designs. In this case, one must view our choice function as resulting from the (statistical) choices of a population of comparable subjects facing different restrictions.\footnote{The literature on stochastic choice also usually adopts this interpretation of a stochastic choice function as resulting from population data.} We illustrate the connection of our work with three existing experiments. The first experiment, a seminal work on reactance by \cite{brehm66social}, exposed subjects to varying strengths of external recommendation to choose between two options. The more forceful the recommendation, the more likely subjects were to choose the unrecommended option.\footnote{Formally, the agents must choose between the recommended option $x$ and the alternative option $y^i$, where $i \in \{low,\ medium,\ high\}$ is the (psychological) cost associated with the choice $y^i$. For instance, $y^{low}$ is the choice of the unrecommended action where the pressure to choose $x$ is \textit{low}. Subjects are more inclined to choose $x$ when the cost of choosing $y$ is \textit{low} than when it is \textit{medium} or \textit{high}.} The second experiment is a classic work by \citet{brehm1977physical}, which investigated whether two-year-old children became attracted to a toy when a physical barrier of varying height impeded access to it. In each condition, a child entered a room where two toys had been placed in opposite corners, and the experimenter recorded the time it took the child to touch each toy. The toys were of two possible types, and the barriers were of three different heights: no fence, a low fence, and a high fence. When the two available toys were of different types and one was behind the high fence, children approached the latter first. However, this strict preference disappeared if there was no fence or if the fence was low. Finally, when only one type of toy was presented, the high fence became repulsive, and children approached the toy without a fence (or behind a low fence) first.\footnote{Let $T_i^j$ denote the toy of type $i \in \{1,2\}$ placed behind the fence of height $j\in \{0,low,high\}$. One can construct a choice correspondence on binary choice as follows: $x$ is chosen in $\{x,y\}$ if the time to get to $x$ is not statistically significantly lower than the time to get to $y$. Except for the presence of indifference, this would fit our setting, and the children exhibit the type of choice reversals that our model aims to capture.} The last experiment is from the economics literature: \cite{AradRubinstein2018} documented a backfire effect of nudges. They found that subjects who were offered different saving programs tended to save more when they received prior information about the benefits of saving than when they were told that a default option had been specifically designed to induce them to save more money. The authors explicitly referred to reactance to explain their results.

\paragraph{The dynamics of the FFE.} 

Since the FFE is often framed as a response to the \textit{elimination} of previous opportunities,  one may contend that this phenomenon is inherently dynamic and should be modeled as such. Yet, at least two arguments justify approaching the FFE with our static framework. First, we previously illustrated that numerous experimental works documented the FFE using between-subject designs that did not involve dynamics. Our approach is therefore better suited to capture the findings of these works than a model incorporating explicit dynamics. Second, our model can be fruitful in exploring interesting dynamic questions about the FFE. A dynamic model would typically study the relation of individuals' propensity to react to restrictions to their previous experiences of freedom and restriction. This relation is far from being unambiguous. Do people only react to the loss of opportunities they previously benefited, or do they become more demanding --- and thus more prone to react to the absence of additional options --- once new freedoms have been granted? Studying the evolution of individuals' attitudes to restriction first requires eliciting these attitudes, which is precisely what our axiomatization exercise accomplishes. Accordingly, rather than challenging the potentially dynamic nature of the FFE, we aim for this paper to serve as a useful first step toward its investigation. More specifically, we believe that our key concepts --- the threshold and reaction preference --- can be fruitfully extended to analyze dynamic theories compatible with the FFE. This is also an opportunity to stress the plurality of such theories, with the aim of encouraging further research along these lines.

Let us frame this discussion more formally. In RSC, the DM's attitudes toward restriction are captured by two ingredients: the thresholds $\{x^\star_T\}_{T\in \mathcal{T}}$ and the reaction preference $\succsim_2$. The DM's prior experiences of restrictions and freedom may shape these ingredients in a non-trivial way. One possible theory holds that people's demand for freedom stems from the opportunities they previously had access to. This hypothesis could imply, for instance, that the thresholds are the best options from each type that were previously available to the DM. Importantly, this theory predicts that the DM only reacts to restrictions that remove options from her opportunity sets. 
An alternative theory, drawing on the \textit{Tocqueville effect}, postulates that experiences of expanding freedom may heighten frustration over remaining constraints and thus prompt new reactions to them.\footnote{See \cite{davies1962toward}.} In our model, this would translate into the requirement that, once the DM is granted the freedom to choose from $A$ rather than the previously available set $B\subseteq A$, she becomes more demanding with respect to options outside $A$ --- for example, by raising the thresholds to options better than those contained in $A$ or by modifying $\succsim_2$. Finally, a third theory might posit that, although the DM initially experiences frustration when her freedom is curtailed, she becomes less demanding as she gets used to the loss. In our framework, this would correspond to the freedom demands embodied by the thresholds and $\succsim_2$ becoming less stringent as restrictions gradually increase.\footnote{We thank a referee for suggesting this hypothesis. In particular, this means that a constraining policy like the integration policy we consider in our application in Section \ref{integration_policy} may be effective in the long-run, even if it backfires in the short run because of the FFE.} We believe our model is helpful for thinking about these different theories and can therefore offer a valuable starting point for formulating and testing them.

\paragraph{No third party.} One may object that restriction-sensitive choices require the involvement of a third party (e.g., a planner) who limits the DM’s available options. While this line of critique offers a promising direction for future research, a large part of the psychological literature on the FFE studies individual reactions without focusing much on the source of the restrictions. Commodity theory does not assume the source of a commodity's scarcity or additional effort costs to acquire it. The source may indeed originate from a third party --- for instance, a company producing a limited edition --- but in such a case, there is no compelling reason to model it explicitly for our purpose. More fundamentally, scarcity can also arise from impersonal forces, such as a sharp rise in prices, an epidemic, or a strike. Although psychological reactance may seem more directly associated with the presence of a third party, it is important to stress that such presence is not a necessary condition in the relevant literature. In their seminal book, \citet[p.4]{brehm2013psychological} wonder about the kind of source that may be perceived as a threat to one's freedom. They answer that ``\textit{any event} that increases the perceived difficulty of having or of not having a potential outcome threatens the exercise of a freedom.'' The previously mentioned experiment by \citet{brehm1977physical} investigates the impact of physical barriers on children's choices of toys. Our model convincingly captures the outcome of this experiment, whereas a model involving a third party would likely face greater difficulty.

\subsection{Related Choice Theory Literature}
\label{section_lit}

As we already emphasized, RSC can be seen as a two-stage procedure where the DM first filters the options according to the operator $d(.)$ and then maximizes $\succsim_2$ over these shortlisted options. Several bounded rationality choice models with this sequential feature have been proposed in the theoretical literature over the past twenty years \cite[see, among others,][for reviews of existing models]{Apesteguia_Ballester_2013, horan_2016, KOPS_2022}. Many of these models consider a quite general first-stage filtering \citep[see][for an overview]{Tyson_2013}; among which are the \textit{choices with limited attention/consideration} \citep{masatlioglu2012revealed, lleras2017more} and the \textit{rationalization procedure} \citep{kalai2002rationalizing, cherepanov2013rationalization, ridout_21}. Another large class of two-step choice models, that we could call \textit{shortlist methods}, consists in sequentially applying two rationales (i.e., binary relations). The initial model by \cite{manzini2007sequentially} \citep[see also][]{dutta_horan_15, KOPS_2022}, called the \textit{rational shortlist methods} (RSM), only requires the two rationales to be asymmetric. Another version, sometimes called the \textit{transitive shortlist methods} (TSM), requires the two rationales to be transitive (and potentially the second one to be complete) \citep{au_kawai_11, horan_2016, lleras2017more}. TSM is not only a specific case of RSM, it is also a specific case of choices with limited attention and rationalization procedure. 

Our model happens to be a specific case of TSM, and thus of the other models mentioned in the previous paragraph. To see this, note that for any $\langle \mathcal{T}, \succsim_1, \succsim_2 \rangle$, the TSM $(\bigcup_{T\in\mathcal{T}}\succsim_{1|T}, \succsim_2)$ (see Definition \ref{tsm_def}) represents the same choice function. Conversely, we show in Appendix \ref{app:TSM} that TSM does not satisfy either \textbf{NRC} or \textbf{SPR}. 

As highlighted in many of the aforementioned papers, there is also a connection with models that rationalize choices as resulting from strategic considerations in a two-stage game. These strategic considerations can come from two distinct individuals, as in the model of rationalisation by game tree by \cite{Xu_Zhou_07} --- which is a distinct model from ours. They can also result from conflicting selves of the same person, as in the planner-doer models by \cite{gul_pesendorfer_05} or \cite{CHANDRASEKHER_18}. These models differ from ours in several significant ways. First, they differ conceptually as they do not seek to represent behaviors related to the FFE. Second, they are axiomatized based on preferences over menus, rather than final choices as in our framework. Finally, their axioms do not allow for deriving the specific structure of RSCs (especially the types and the single-peakedness of $\succsim_2$ with respect to $\succsim_1$).

The attraction and the compromise effects, which have been widely observed and studied, are also related. While many of the aforementioned papers can accommodate these phenomena, the work by \cite{ok2015revealed} stands apart as it specifically focuses on them and does not belong to any of these classes of models. Their definition of revealed reference is based on choice patterns similar to those exhibited in our definition of $\mathbf{R}^c$, that is, $z=c\{x, y,z\}$ and $x=c\{x,z\}$. Their interpretation is, however, significantly different: they argue that $z$ beats $x$ only with the ``help'' of $y$. Hence, while they interpret these reversals as revealing a relationship between $y$ and $z$, we interpret it as revealing a relationship between $x$ and $y$.\footnote{More precisely, the application of their definition identifies $y$ as a \textit{revealed reference} of $z$.} Their model and ours are actually incompatible. Indeed, as we previously noted, observing $x\mathbf{R}^c y$ and satisfying \textbf{Exp} imply a binary choice cycle, which is prevented by their \textit{No-Cycle Condition}.

Finally, the forbidden fruit effect is related to the literature on phantom alternatives. Several experiments have demonstrated that the presence of this type of alternative, which is desired but unavailable, prompts agents to choose a similar available alternative \citep{farquhar1993decision,pettibone2000examining,highhouse1996context}. In line with this notion, \cite{guney2018aspiration} characterize a model of aspiration-based choice in which the analyst can observe the DM choices over pairs of menus: one comprising feasible alternatives and another comprising potential alternatives, which are not necessarily feasible. In our setting, we do not assume that the analyst observes potential alternatives envisioned by the DM, as is often the case in many applications. The revelation exercise is thus complicated by the absence of such observations.

\bigbreak


\section{Normative analysis}\label{normative}

What are the normative implications of our model? Our theory has original implications on welfare and the assessment of freedom. We address these issues separately in this section.

\subsection{Measure of Welfare}\label{subsec_welfare}

As extensively discussed in the literature, menu-dependent choices pose a serious challenge to  welfare analysis \citep{Bernheim_rangel_07, bernheim2008choice, bernheim_rangel_09QJE, manzini_mariotti_2012, CHAMBERS_2012, rubinstein_salant_12, Apesteguia_ballester_15, grune_yanoff_22}. It is therefore natural to wonder how welfare is to be analyzed in restriction-sensitive choices. Our analysis is model-based \citep{manzini_mariotti_welfare_14}.\footnote{This is in contrast to the so-called ``model-free'' approach proposed by \cite{Bernheim_rangel_07,bernheim2008choice,bernheim_rangel_09QJE}. Any welfare criterion that would result from an RSC will heavily depend on our interpretation of the ingredients of RS-structures.} We interpret $\succsim_1$ as the appropriate welfare criterion for the individual, as it is supposed to represent their preferences without any restriction-related motives. However, not every part of $\succsim_1$ is uniquely identified from choice data. This is why we focus on minimal single-peaked RS-structures (Proposition \ref{min_RS_structure}).\footnote{The focus on minimal single-peaked RS-structure is justified by the fact that the types and the thresholds are essentially unique: in particular, some options $x,y$ might otherwise be mis-categorized as belonging to the same type while $x\not\approx y$ and some conclusions about welfare could be wrongly drawn.} Among minimal single-peaked RS-structures, $\succsim_1$ is uniquely identified only for options that lie in the same type or for options that are above the threshold $x^\star_T$ of their respective type --- more precisely, the transitive closure of these identified relations is unique. Hence, we propose the following definition.\footnote{Again, $\mathcal{T}(x)$ denotes the type of option $x$.}

\begin{definition}
    Suppose $c$ is a single-peaked RSC and let $\mathcal{S}= \langle \mathcal{T},  \succsim_1,\succsim_2 \rangle$ be a minimal single-peaked RS-structure that rationalizes $c$. For any $x,y \in X$, $x\neq y$, $x$ is \emph{\textbf{welfare improving}} on $y$, denoted $x \gg^c_\mathcal{S} y$, if either:
    \begin{enumerate}[(i)]
        \item $\mathcal{T}(x) = \mathcal{T}(y)$ and $x\succ_1y$; or
        \item $\mathcal{T}(x) \neq \mathcal{T}(y)$, $x\succ_1x^\star_{\mathcal{T}(x)}$ and there exists $z \in \mathcal{T}(y)$ such that $z\succ_1x^\star_{\mathcal{T}(y)}$, $x\succ_2z$, and $ z\succ_1y$.
    \end{enumerate}
\end{definition}

Note that it easily follows from Proposition \ref{min_RS_structure} that two distinct minimal single-peaked RS-structures generate the same welfare improving relation. We thus simply write $\gg^c$, where $\gg^c \ = \ \gg^c_\mathcal{S}$ for any minimal single-peaked RS-structure $\mathcal{S}$ that rationalizes $c$.

Interestingly, we show by means of examples that neither the conservative criterion $P^\star$ of \cite{bernheim_rangel_09QJE} nor the preference identified from \textit{choice with limited attention} (CLA) $P^R$ \citep{masatlioglu2012revealed} coincide with our criterion\footnote{$P^\star$ is defined as follows: for any $x$ and $y$, $xP^\star y$ if $x$ is sometimes chosen when $y$ is feasible while $y$ is never chosen when $x$ is feasible. $P^R$ is the transitive closure of the relation $P$ defined in the following way: for any $x$ and $y$, $xPy$ if there exists $A$ such that $x=c(A) \neq c(A\setminus\{y\})$.} $\gg^c$. The latter point is worth emphasizing as RSC is a special case of CLA (see Section \ref{section_lit}).

\paragraph{Example 1: $\gg^c \ \not\subset P^\star$.} Let $z=c\{x,y,z\}$ and $x=c\{x,z\}$ and suppose that $z\succ_1x^\star_{\mathcal{T}(z)}$, $y\succ_1x^\star_{\mathcal{T}(y)}$. Then $z \gg^c x$ while $x=c\{x,z\}$, hence $\neg[zP^\star x]$.

\paragraph{Example 2: $P^\star \not\subset \ \gg^c$.} Consider $x$ and $z$ such that $x^\star_{\mathcal{T}(x)}\succ_1 x$, $z \notin \mathcal{T}(x)$, and for every $y \in \mathcal{T}(x)$, $y\succ_2z$. Therefore, whenever $x$ is available, $z$ is never chosen and $x = c\{x,z\}$; hence $x P^\star z$, while $\neg[x \gg^c z]$.

\paragraph{Example 3: $\gg^c \ \not\subset P^R$ and $P^R \not\subset \ \gg^c$.} This is inspired from the example used in \cite{masatlioglu2012revealed} to demonstrate the difference between $P^R$ and $P^\star$ (Example 1, pp. 2191-2192). $X=\{x,y,z,t\}$, there are two types $\{x,y\}$ and $\{t,z\}$, $y$ and $z$ are above their respective thresholds while $x$ and $t$ are below, $t\succ_2x \succ_2 z\succ_2 y$. Therefore, the application of \cite{masatlioglu2012revealed}'s relation $P$ would imply that $xPz$ and $zPy$, hence $xP^R y$; whereas we would conclude that $y\gg^c x$.

\subsection{Measure of Freedom}\label{subsec_freedom}

Political philosophers widely agree that normative evaluations that focus on welfare analysis neglect the importance of considering freedom concerns \citep{rawls2020theory,nozick1974anarchy,dworkin2000sovereign}. Starting with \cite{jones_sugden82}, \cite{Sen88}, \cite{pattanaikxu90}, and \cite{Puppe1996}, economists have tried to cope with this issue, and a wide variety of freedom measures --- mostly based on the opportunity sets --- have been proposed \citep[see][for a survey of this literature]{Baujard07}. These attempts usually adopt an objective stance by assuming that the DM's subjective viewpoint on their freedom is irrelevant to the evaluation. Under the reactance interpretation of our model, RSC allows for incorporating a subjective dimension in the evaluation of the DM's freedom. 
 
In our view, which is consistent with the psychological reactance theory, each type embodies a specific freedom. We focus on minimal single-peaked RS-structures. A freedom $T$ is \textbf{satisfied} whenever the opportunity set contains at least one option above the threshold, that is, when the opportunity set intersects with the set
\begin{align*}
    F^T\equiv \{x\in T: x\succ_1 x^\star_{T}\}.
\end{align*}

Our measure simply counts, for each opportunity set, the number of satisfied freedoms.\footnote{Importantly, this measure combines two dimensions that have been separately pointed out as relevant to the agents' valuation of their freedom in the literature: their (potential) preferences over options \citep{Sen93, pattanaik1998preference, Puppe1996}; and the similarity between different options \citep{pattanaik2000diversity,nehring2002theory}. Preferences are integrated through the fact that only sufficiently good options (i.e., above $x^\star_T$) matter for the DM's evaluation, and the similarity between options is captured through the types.} We characterize this measure with two axioms. Let $\langle \mathcal{T}, \succsim_1, \succsim_2 \rangle$ be a minimal single-peaked RS-structure defined on $X$, and $\succsim$ a complete and transitive binary relation defined on $\mathcal{X}$.

A menu $A$ is \textbf{richer than} a menu $B$ if for any $T\in\mathcal{T}$, if $ F^T \cap A = \emptyset$, then $ F^T \cap B = \emptyset$. Furthermore, we say that $A$ is \textbf{strictly richer than} $B$ if $A$ is richer than $B$ but the reverse is not true. Our first axiom naturally requires that (strictly) richer sets be always (strictly) preferred, and imposes that it be an equivalence for singletons.

\paragraph{\textsc{R-Dominance}.}
\textit{ 
\begin{enumerate}[(i)]
    \item For any $A,B \in \mathcal{X}$: if $A$ (strictly) richer than $B$, then $ A (\succ) \succsim B$;
    \item For any $x,y \in X$: $\{x\} \succ \{y\} \implies$ $\{x\}$ strictly richer than\footnote{Note that part (i) of the axiom implies monotonicity in the sense of \cite{kreps1979representation}: for any $A, B \in \mathcal{X}$, $A \supseteq B \implies A \succsim B$. Indeed, in this case, $A$ is trivially richer than $B$. Part (ii) is an adaptation of \cite{pattanaikxu90}'s \textit{Indifference Between no Choice Situations}, which simply imposes an indifference between every singleton. They argue that singletons do not offer any freedom of choice, hence they cannot be strictly ranked. This is still true in our case, except if only one the two options is above the threshold of its type (i.e., in $F$), which is exactly what is implied by (ii).} $\{y\}$.
\end{enumerate}
}

\vspace{0.3cm}

Our second axiom is an adaptation of the composition axioms used in Pattanaik and Xu's series of papers, which states that combining menus that do not overlap should preserve the ranking. In our approach, this must however be true only if combining really provides additional freedom, that is, when you combine opportunity sets that are not richer than one another.

\paragraph{\textsc{R-Composition}.}
\textit{For any $A, B, C, D \in \mathcal{X}$, such that $A\cap C = B\cap D = \emptyset$, $C\subseteq T$ and $D\subseteq T'$ for some $T,T' \in \mathcal{T}$, and $A$ is not richer than $C$: if $A \succsim B$ and $C\succsim D$, then $A\cup C \succsim B \cup D$.
}

\vspace{0.5cm}

For any menu $A$, we define $n(A)$ as the number of satisfied freedoms in $A$, that is, the number of type $T\in\mathcal{T}$ such that  $F^T\cap A\not=\emptyset$. We can now state our representation theorem.

\begin{theorem}\label{menu_pref}
$\succsim$ satisfies \textsc{R-Dominance} and \textsc{R-Composition} iff for any menu $A$ and $B$:
$$A \succsim B \iff n(A) \geq n(B)$$
\end{theorem}


\section{Applications}\label{applications}

We explore the scope of applicability of RSC. We show how our model can plausibly explain two empirically supported phenomena: the backfire effect of beliefs and the backlash of integration policies targeted toward minorities. Hereafter, we adopt the interpretation of our model and the vocabulary in terms of reactance.

\subsection{Backfire Effect of Beliefs}

As \cite{sensenig1968attitude} suggest, reactance has its counterpart in the realm of beliefs; namely, the boomerang effect for psychologists \citep{hovland1953communication}, or the backfire effect for political scientists \citep{nyhan2010corrections,wood2019elusive}.\footnote{The boomerang effect is ``a situation in which a persuasive message produces attitude change in the direction opposite to that intended.'' The backfire effect is a concept from political science that refers to a situation in which evidence contradicting the subjects' prior belief may reinforce their belief in the opposite direction.} We propose to accommodate this mechanism by adapting \citeauthor{che_mierendorff19}'s \citeyearpar{che_mierendorff19} single period model of attention allocation with reactance.

A DM must choose from two actions, $l$ or $r$, whose payoffs depend on an unknown state $i \in \{L,R\}$. His prior belief that the state is $R$ is denoted $p$ and we assume that $p \in (0,1/2)$. Before choosing his action, the DM can acquire information by allocating his attention across four sources of information (e.g., newspapers). 

The sources are represented by statistical experiments. For $i=L,R$, there are two $i$-biased experiments, denoted $\sigma^{ii}$ and $\sigma^i$, that can only reveal the state $-i$. $\sigma^{ii}$ is an \textit{extreme} source, whereas $\sigma^i$ is a \textit{moderate} one, that is, the former is more biased than the latter. Formally, $\sigma^i$ sends signal $s^i$ with probability $1$ in state $i$ and with probability $1-\lambda$ in state $-i$, and $\sigma^{ii}$ sends signal $s^i$ with probability $1$ in state $i$ and with probability $1-\delta$ in state $-i$. We assume that $3/4 > \lambda > \delta =1/2$. The experiments induced by the moderate sources $\sigma^L$ and $\sigma^R$ are described in Table \ref{moderate_exp}. The signals $\sigma^{LL}$ and $\sigma^{RR}$ are obtained by replacing $\lambda$ with $\delta$.

\begin{table}[htb]
    \centering
    \hfill
    \begin{tabular}{c|cc}
        \multicolumn{3}{c}{} \\
        \multicolumn{3}{c}{$\sigma^L$} \\
        \hline
        \hline
           State/signal    & $s^L$ & $s^R$ \tabularnewline
        \hline
        $L$ & $1$ & $0$ \tabularnewline
        $R$ & $1-\lambda$ & $\lambda$
        \tabularnewline
    \end{tabular}
    \hfill
 \begin{tabular}{c|cc}
        \multicolumn{3}{c}{} \\
        \multicolumn{3}{c}{$\sigma^R$} \\
        \hline
        \hline
          State/signal     & $s^L$ & $s^R$ \tabularnewline
        \hline
        $L$ & $\lambda$ & $1-\lambda$ \tabularnewline
        $R$ & $0$ & $1$
        \tabularnewline
    \end{tabular}
    \hfill\null
    \caption{Experiments induced by the moderate sources.}
    \label{moderate_exp}
\end{table} 

We assume that the DM's choices of sources of information are made according to the following RS-structure. The sets of $L$-biased sources and $R$-biased sources each represent a type. The welfare preference $\succsim_1$ is represented by the function $u$. The expected value of $u$ is computed from the payoffs $\pi^i_a$ of choosing action $a \in \{l,r\}$ in state $i \in \{L,R\}$, given by: $\pi^R_r=\pi^L_l =1$, $\pi^R_l=\pi^L_r=-1$. Hence, given that for $i=L,R$, $\sigma^i$ is strictly more Blackwell informative than $\sigma^{ii}$, the DM will never choose an extreme source if the associated moderate one is available, that is, $\sigma^{i}\succ_1 \sigma^{ii}$. 

The reaction preference $\succsim_2$ is represented by the function $v$. The expected value $v$ equals that of $u$ for moderate sources of information $\sigma^L$ and $\sigma^R$. For the extreme source $\sigma^{RR}$, $v$'s expectation is computed from the payoffs $\hat\pi^L_r=0$, the other payoffs being the same. Similarly, for the extreme source $\sigma^{LL}$, $v$'s expectation is computed from the payoffs $\hat\pi^R_l=0$, the other payoffs being the same. In other words, the absence of a moderate source $\sigma^i$ from a menu makes the DM change her assessment of a mistake made in state $i$, increasing the value of extreme sources.

Suppose first that the DM has access to the large menu $M=\{\sigma^{LL},\sigma^L,\sigma^R,\sigma^{RR}\}$. In this case, $d(M)=\{\sigma^L, \sigma^R\}$, and she prefers action $r$ if and only if her posterior belief is greater than $1/2$. One can show that she optimally allocates her attention by choosing the ``own-biased news source'', that is, the signal biased toward her prior: in our case, $\sigma^L$, given that $p < 1/2$. The rationale for this is the following. The prior indicates action $l$ as the optimal one. Hence, a breakthrough signal $s^R$ from $\sigma^L$ is more valuable than a breakthrough signal $s^L$ from $\sigma^R$. And the biased signal $s^L$ from $\sigma^L$ is more aligned with the DM's prior belief than $s^R$ from $\sigma^R$. Hence, she is better off allocating her attention to $\sigma^L$ \cite[see][pp. 2999-3000, for the complete argument]{che_mierendorff19}. 

Suppose now that the moderate R-biased source $\sigma^R$ is no longer available --- e.g., because the government banned this newspaper. The DM only has access to $L$-biased or extremely $R$-biased sources, the menu $N=\{\sigma^{LL},\sigma^L,\sigma^{RR}\}$. $\sigma^{RR}$ is no longer removed from consideration by $\sigma^R$, hence $d(N)=\{\sigma^L,\sigma^{RR}\}$. Her utility from choosing $\sigma^L$ is unchanged while the one attached to $\sigma^{RR}$ is $p + (1-p)\delta$ (for $p$ sufficiently close to $1/2$ such that after signal $s^R$ from $\sigma^{RR}$, the DM chooses action $r$). As a consequence, some DMs with prior beliefs sufficiently close to $1/2$, who would have chosen news source $\sigma^L$ in menu $M$, will choose the extreme source $\sigma^{RR}$ in menu $N$ and their default action becomes $r$.

\begin{proposition}\label{conspiracy}
There exists $p^\star<1/2$ such that if $p \in [p^\star, 1/2]$:
\begin{enumerate}[(i)]
    \item The DM prefers $\sigma^{RR}$ to $\sigma^L$ in menu $N$;
    \item After a realisation of signal $s^R$ from $\sigma^{RR}$, the DM chooses action $r$.\footnote{All proofs of this section can be found in Appendix \ref{proof_appli}.} 
\end{enumerate} 
\end{proposition}

This is in strong opposition to what would be obtained without reactance. Indeed, if the DM does not modify her utility when the menu shrinks, by removing $\sigma^R$, some DMs with prior belief strictly higher than $1/2$ would now choose the source $\sigma^L$ instead and action $l$ after a signal $s^L$.

One implication is that an $R$-biased news source seeking to attract more consumers would have an interest in adopting a strong bias and making it credible that there are no moderately $R$-biased sources available. Similarly for an $L$-biased news source. Therefore, suppose that two oppositely biased news sources compete over a mass one of consumers with heterogeneous priors and decide for that purpose whether to adopt a moderate or an extreme position. This competition would lead to the polarization of the news sources; that is, both competitors would propose extremely biased content.

\subsection{Integration Policy Backlash}\label{integration_policy}

Can forced assimilation policies foster the integration of minorities? While \citeauthor{alesina2015nation}'s \citeyearpar{alesina2015nation} theory of nation-building assumes that repressing the cultural practices of minorities spurs homogeneity, \citet{bisin2001economics} suggest that the success of such policies may be mitigated by an increasing effort of parents to influence their children's cultural traits. In this application, we show that, with reactance, one can even predict this policy to yield a backlash effect: the repressed minorities react to repression by becoming more prone to self-isolation. This mechanism additionally provides a rationale for the persistence of reactance as an evolutionarily efficient behavior.

Such a backlash effect has been empirically documented. The evidence suggests that the ``burkha ban'' in France in 2004 has strengthened the religious identity of French-Muslims \citep{abdelgadir_fouka_2020}. \citet{fouka2020backlash} shows that, in states which prohibited German Schools in the aftermath of World War I, German-Americans ``were less likely to volunteer in World War II and more likely to marry within their ethnic group and to choose decidedly German names for their offspring.'' 

To show how this backlash operates, we complement \citeauthor{bisin2001economics}'s \citeyearpar{bisin2001economics} account of cultural transmission with a reactance mechanism: as repression increases, parents' educational freedom decreases and, reacting to this repression, they may endeavour to influence their children even more.\footnote{For simplicity, we adopt a continuous setting, while our own framework is discrete. The ideas would be exactly the same with a discrete setting.} There are two cultural traits $\{m,M\}$ --- for $m$inority and $M$ajority. The proportion of the minority $q$ is assumed to be positive but lower than $1/2$. From now on, we focus on the minority; the same mechanism applies to the majority. Each generation is composed of parents who have only one child. Intergenerational transmission results from two socialization mechanisms. By vertical socialization, the parents may directly transmit their cultural trait $m$ with probability $d$. If, with probability  $1-d$, vertical socialization fails, then horizontal transmission occurs and the child adopts the traits of a random individual in society. Hence, the probability that a child from the minority is socialized by her parents' traits is:
\begin{align}
    P(d) \equiv d +(1-d)q. \label{socialization}
\end{align}

As \citet{bisin2001economics}, we argue that parents endeavor to influence their child. They have a unit of time to allocate between their effort to fix $d$ --- which costs $d^\beta$ unit of time, with $\beta>1$ --- and a leisure activity $t \in [0,1]$, whose cost and utility are $t$. In addition, the government can implement a repressive policy $g\geq 1$ that may increase the parents' cost of influencing their child: a pair $(t,d)$ costs $t+d^\beta g$ units of time for the parents. We posit that parents get a utility of 0 when their child is socialized to the other trait, while they get a utility $V(g)$ when she is socialized to their own trait. Hence, the expected utility of their child's socialization is $P(d)V(g)$. This means that, given a repressive policy, parents choose options $(t,d)$ from the menu 
\begin{align*}
    K_{g}\equiv \{(t,d) \in [0,1]^2 : t+d^\beta g\leq 1\},
\end{align*}
to maximize
\begin{align}
    t+P(d)V(g),
\label{program35}
\end{align}
We assume that $V$ has the following shape:
\begin{align*}
    V(g)= \left\{\begin{array}{cc}
        \hat{V} & \text{ if }\hat{g}\geq g \\
        \hat{V}(\frac{g}{\hat{g}})^{\lambda}& \text{ if } \hat{g}< g
    \end{array}\right.
\end{align*}
for some $\hat{g}>1$ with $\lambda>1$ and $\hat{V}>1$. Hence, above the threshold $\hat{g}$, the more repressive is the policy $g$, the greater is $V(g)$. The interpretation is that parents react to the repressive policy when they feel that their freedom to educate their child is threatened. In other words, more repression may create incentives to dedicate more resources to transmit their traits to their children. Note that $\lambda$ represents some kind of reactance rate since, as it increases, parents' willingness to influence their child also increases. 

From the first order condition, we obtain that the unique equilibrium educational effort --- the program (\ref{program35}) being concave --- must satisfy:
\begin{align}
    d^{m\star}(g,q) =\min \Bigg\{\bigg(\frac{1}{g}\bigg)^{\frac{1}{\beta}},\bigg(\frac{1-q}{\beta}\frac{V(g)}{g}\bigg)^{\frac{1}{\beta-1}}\Bigg\}
    \label{FOC}
\end{align}

Let us assume that $d^{m\star}(.,q)$ is not a corner solution at $\hat{g}$, that is, $d^{m\star}(\hat{g},q)<(1/\hat{g})^{1/\beta}$. Let $\Bar{g}>\hat{g}$ be the lowest threshold such that for any $g\geq \Bar{g}$, $d^{m\star}(g,q)=(1/g)^{1/\beta}$. Then given the shape of $V$, $d^{m\star}$ decreases with $g$ on $[1,\hat{g}]$, increases on $(\hat{g},\Bar{g}]$, and decreases on $(\bar{g}, +\infty)$. In other words, when the repressive policy is in $[\hat{g}, \Bar{g}]$, the more repression, the more parents invest in having their child socialized to their own trait. This suggests that reactance is at work in this model. In the following lemma, we establish the precise connection between this model and RSC. 
\begin{lemma}\label{Connec}
    The function $c$ defined on $\{K_{g}\}_{g \geq 1}$, such that for all $g$
    \begin{align*}
        c(K_{g})=\{(t,d)\in K_{g} : (t,d)\text{ solves }(\ref{program35})\}.
    \end{align*}
    is a well-defined choice function and there exists an RSC $c'$ defined on all compact subsets of $[0,1]^2$ such that $c(K_g)=c'(K_g)$ for all\footnote{For convenience, we construct an RS-structure on this infinite collection of compact sets. Obviously, analogous results could be obtained by making the set of possible policies $g$ and the menus $K_g$ finite.} $g\geq 1$.
\end{lemma}

Assuming the repressive policy solely concerns the minority (i.e., there exist $g^m>1$ and $g^M=1$), what does reactance imply for the population dynamics in this model? Let time $\tau \in [0, +\infty)$ be continuous and $q_\tau$ be the share of the population with the minority cultural trait at time $\tau$. Then, we have\footnote{See \citet[eq. (3), footnote 9]{bisin2001economics} for discussions about this differential equation.}
\begin{align*}
    \Dot{q}=q(1-q)\Big(d^{m\star}\big(g^m,q\big)-d^{M\star}\big(1,1-q\big)\Big).
\end{align*}

Assume that $g\in [\hat{g},\bar{g}]$. Then by (\ref{FOC}), $d^{m\star}$ satisfies the \textit{cultural substitution property}.\footnote{In \citet[Definition 1]{bisin2001economics}, this property states that $d$ is continuous, decreasing with $q$, and $d=0$ when $q=1$.} This implies that $q$ converges to some $q^\star\in (0,1)$, which satisfies $d^{m\star}\big(g^m,q\big)=d^{M\star}\big(1,1-q\big)$ \citep[see][Proposition 1]{bisin2001economics}. Hence, 
\begin{align}
    q^\star(g^m)=\frac{V(g^m)/g^m}{V(1)+V(g^m)/g^m}.
    \label{dydy}
\end{align}
Given that $V(g)/g$ increases with $g$ when $g\in [\hat{g}, +\infty)$ this means that, above $\hat{g}$, repressive policy increases the size of the minority. This prediction contrasts with \citeauthor{alesina2015nation}'s \citeyearpar{alesina2015nation} suggestions. 

Noting that reactance is presumably a characteristic cultural trait \citep{jonas2009culture}, this model also provides a rationale for why reactance can be evolutionary efficient. Minorities that are more prone to exhibit reactance are more likely to survive repressive attempts to hinder their cultural practices. To make precise this comparative statics statement, consider two minorities: one with a high reactance rate $\lambda^H$ and one with a low reactance rate $\lambda^L<\lambda^H$. Denoting  by $q^\star_L(.)$ and $q^\star_H(.)$
the equilibrium population share for these two minorities, the following proposition establishes that $q^\star$ is always higher for the high-reactance minority.
\begin{proposition} 
For all $g>\hat{g}$, $q^\star_H(g)>q^\star_L(g)$.
\label{dydy2}
\end{proposition}


\section*{Appendices}
\begin{appendices}
\sectionfont{\centering\large}

\section{Proofs of Section \ref{section:rsc}}

All along the proofs, for an RS-structure $\langle \mathcal{T},  \succsim_1, \succsim_2 \rangle$, we denote as $\mathcal{T}(x)$ the type to which belongs the option $x$. Let us define the binary relation $\succsim \subset X^2$ by $x\succsim y$ if and only if $x=c\{x,y\}$ or $x=y$ and denote by $\succ$ its asymmetric part. 

\subsection{Proof of the Propositions and Supporting Lemmas}

\begin{proof}[\textbf{Proof of Proposition} \ref{RSC_property}]
Let $c$ be an RSC rationalized by $\langle \mathcal{T},  \succ_1, \succ_2 \rangle$. The \textit{if} part is left to the reader as it simply results from an application of the choice procedure to the sets $\{x,y,z\}$ and $\{x,z\}$. We prove the \textit{only if} part. Consider $x,y,z$ such that $z=c\{x,y,z\}$ and $x = c\{x,z\}$, so $x \mathbf{R}^c y$. One can easily check that it is not possible that $x,y,z$ are either all in the same type, or all in different types: in both cases, the choices between any combination of these options result from the maximization of a unique order, so it must satisfy Property $\alpha$. Therefore, exactly two among them must be of the same type, denoting it $T$. Given that $z=c\{x,y,z\}$, this means that $z\in d\{x,y,z\}$ and $z$ is the best element of its type according to $\succ_1$. Consequently, $x$ is not of the same type as $z$ as otherwise this would imply that $z=c\{x,z\}$. This also implies that $z\in d\{x,z\}$ and thus $x\succ_2 z$. Therefore, $x\notin d\{x,y,z\}$, as otherwise it would imply $x = c\{x,y,z\}$, which is only possible if $x \in \mathcal{T}(y)$ and $y\succ_1 x$. Therefore, $y\in d\{x,y,z\}$ and $z=c\{x,y,z\}$ imply that $z \succ_2y$.
\end{proof}

\begin{lemma} 
    Assume $c$ is rationalized by an RS-structure $\langle \mathcal{T}, \succsim_1, \succsim_2 \rangle$. For each $T'\in X_\approx$, there exists $T\in \mathcal{T}$ such that $T'\subseteq T$.
    \label{IncluTT}
\end{lemma}
\begin{proof}
    Let $ x,y\in T'\in X_\approx$, i.e. $x\approx y$. Given that $X$ is finite, the definition of $\approx$ implies the existence of $n$ and $x_1,\dots ,x_n\in X$ such that  $(x\mathbf{R}^cx_1\vee x_1\mathbf{R}^cx)\wedge \dots\wedge (y\mathbf{R}^cx_n \vee x_n\mathbf{R}^cy)$. By Proposition \ref{RSC_property}, for all $k<n$ $(x_k\mathbf{R}^cx_{k-1} \vee x_{k-1}\mathbf{R}^cx_k)$ implies $\mathcal{T}(x_{k-1})=\mathcal{T}(x_{k})$. Hence, by induction we obtain $\mathcal{T}(x)=\mathcal{T}(x_1)=\dots =\mathcal{T}(x_{n})=\mathcal{T}(y)$, proving $T'\subseteq \mathcal{T}(x)\in \mathcal{T}$.  
\end{proof}

\begin{lemma}\label{interemediarytypes}
    Assume $c$ is rationalized by an RSC $\langle \mathcal{T}, \succsim_1, \succsim_2 \rangle.$ For any $T\in \mathcal{T}, T'\in X_{\approx}$ with $T\cap T'\not=\emptyset$ and any $x\in T\setminus T'$, either $x=c\{x, y\}$ (and thus $x\succ_1y$) for any $y\in T'\cap T$ or $y=c\{x, y\}$ (and thus $y\succ_1x$) for any $y\in T'\cap T$.
\end{lemma}
\begin{proof} Note first that if $|T'|=1$, there is nothing to prove. If $|T'|>1$, by way of contradiction, assume that there exist $y,z\in T'$ such that $y=c\{x,y\}$ and $x=c\{x,z\}$. We thus have $y\approx z$, $y\succ_1x\succ_1 z$, and the sets $U(x)=\{y'\in T': y'\succ_1 x\}$ and $L(x)=\{y'\in T': x\succ_1 y'\}$ are nonempty. Moreover, by Lemma \ref{IncluTT}, $U(x)\cup L(x)\subseteq T$. Assume for all $y'\in U(x)$ there exists no $z'\in L(x)$ such that $z'\mathbf{R}^cy'$. Since for all $(y',z')\in U(x)\times L(x)$, $y'\succ_1z'$, Proposition \ref{RSC_property} entails also $\neg[y'\mathbf{R}^cz']$, and therefore $z'\not\approx y'$, which contradicts $z\approx y$. Hence, we can take $y'\in U(x)$ and $z'\in T'$ such that $z'\mathbf{R}^cy'$. By Proposition \ref{RSC_property}, $y'\succ_1 z'$ and $z'\succ_2 t\succ_2 y'$ for some $t\notin T$. If $x\succ_1z'$, then by Proposition \ref{RSC_property} either $x\succ_2t$ and so $x\mathbf{R}^cy'$ or $t\succ_2x$ and so $z'\mathbf{R}^cx$. Both scenarios entails by the transitivity of $\approx$ that $x\approx u$ for some $u\in T'$, which contradicts $x\not\in T'$. Hence, $z'\succ_1x$. This proves that for all $y'\in U(x)$, $z'\mathbf{R}^cy'$, implies $z'\in U(x).$ The same reasoning shows that for all $z'\in L(x)$, $z'\mathbf{R}^cy'$ implies $y'\in L(x).$ Hence, for all $(y',z')\in U(x)\times L(x)$, $y'\succ_1z'$, and thus $\neg[y'\mathbf{R}^cz']$. It follows that for all $y'\in U(x), z'\in L(x)$, $y'\not\approx z'$. In particular, $y\not\approx z$. A contradiction. 
\end{proof}

\begin{proof}[\textbf{Proof of Proposition} \ref{proposition:types}] 
    This is follows the proof of Theorem \ref{thm1}, in which we construct an RS-structure that satisfies that $\mathcal{T}=X_\approx$. The rest of the statement follows from Lemma \ref{IncluTT}. 
\end{proof}

\begin{proof}[\textbf{Proof of Proposition} \ref{min_RS_structure}] 
    This follows the proof of Theorems \ref{thm1} and \ref{THMSinglePeaked}, in which we construct an RS-structure and show that it satisfies this property. 
\end{proof}
\subsection{Proof of Theorem \ref{thm1}}
 
\begin{proof}[Necessity of the axioms]
    Let $c$ be an RSC rationalized by $\langle \mathcal{T}, \succsim_1, \succsim_2 \rangle$.

    \textbf{Exp}. Let $x \in X$ and $A, B \in \mathcal{X}$ be such that $x = c(A) = c(B)$. This means that $x\in d(A)\cap d(B)$. Hence, $x\succ_1y$ for all $y\in (A\cup B)\cap \mathcal{T}(x)$, $y \neq x$, which implies that $x\in d(A\cup B)$. Moreover, $x=c(A)=c(B)$ implies that $x\succ_2z$ for all $z\in d(A)\cup d(B)$. Besides,  $d(A\cup B) \subseteq d(A)\cup d(B)$, hence $x\succ_2z$ for all $z\in d(A\cup B)$. Hence, $x =c(A\cup B)$.

    \textbf{NRC}. Let $x,y,z\in X$ be such that $x\approx y \approx z$. By Lemma \ref{IncluTT}, $\mathcal{T}(x)=\mathcal{T}(y)=\mathcal{T}(z)$. Hence, $x\succ y\succ z$ is equivalent to $x\succ_1 y\succ_1 z$, which by the transitivity of $\succ_1$ implies  $x\succ_1 z$, and thus $x\succ z$. 

    \textbf{IR}. Let $x,y,z,t\in X$ such that $x\approx y\not\approx z,t$, $x\succ z\succ y$, and $y\succ t$. By Lemmas \ref{IncluTT} and \ref{interemediarytypes}, $\mathcal{T}(x)=\mathcal{T}(y)\neq\mathcal{T}(z)$. Hence, $x\succ z\succ y$, stands for $x\succ_2 z\succ_2 y$, i.e by transitivity $x\succ_2y$. Since $t\not \approx x$ either $t\notin \mathcal{T}(x)$ or $t\in \mathcal{T}(x)\setminus T$ where $T\in X_{\approx}$ and $x,y\in T$. The first case implies $y\succ_2 t$, and the transitivity of $\succ_2$ implies $x\succ_2t$ and thus $x\succ t$, as desired. By Lemma \ref{interemediarytypes}, in the second case we have that $y\succ t$ implies $x\succ t$, as desired.
\end{proof}

\begin{proof}[Sufficiency of the axioms]
    Let $\mathcal{T}=X_{\approx}$. We define, for any $T\in \mathcal{T}$, the relation $\unrhd_T\subseteq T^2$ as follows. For any $x,y \in T$,
    \begin{align}\label{KeyRelation3}
        x\unrhd_T y \iff \big[\forall z\notin T, \quad y\succ z\implies x\succ z\big].
    \end{align}
    We denote respectively as $\rhd_T$ and $\bowtie_T$ its asymmetric and symmetric parts. Note that $\unrhd_T$ is transitive for all $T\in \mathcal{T}$, given that \textbf{NRC} implies that $\succ$ is transitive within type. $\unrhd_T$ is reflexive, and thus by \textbf{IR}, it is complete. The following lemma characterizes $\mathbf{R}^c$ with the relations $\unrhd_T$ and $\succsim$.

\begin{lemma}\label{Lemmarelou2}
    For all $x,y\in T  \in X_\approx$, $x\mathbf{R}^cy$ is equivalent to $x\rhd_{T}y$ and $y\succ x$.
\end{lemma}
\begin{proof}
     $x\mathbf{R}^cy$ means that there exists $z\in X$ such that $x\succ z$ and $z=c\{x,y,z\}$. By \textbf{Exp}, this implies that $y\succ x\succ z\succ y$. Since $x\approx y$, \textbf{NRC} implies that $z\notin T$. Hence, $y\succ x\succ z\succ y$ implies $\neg[y\unrhd_Tx]$ and $y\succ x$, which by completeness of $\rhd_T$ amounts to $x\rhd_Tz$ and $y\succ x$. 

     Conversely, $x\rhd_T y$ and $y \succ x$ imply that there exists $z \notin T$ such that $x\succ z \succ y \succ x$. If $c\{x,y,z\} \in \{x,y\}$, then either $y\mathbf{R}^c z$ or $z\mathbf{R}^cx$, both of which contradict that $z\notin T$. Hence, it must be that $z=c\{x,y,z\}$, implying that $x\mathbf{R}^c y$.
\end{proof}

By \textbf{NRC}, $\succsim$ is a linear order when restricted to any $T\in \mathcal{T}$. Hence, the following option, for a given $T$, is well-defined: $x^T = \min\big(\max(T,\unrhd_T\big), \succsim)$.

We now prove that for each $T\in \mathcal{T}$, the set $\{x \in T \mid \nexists y, x\mathbf{R}^c y\}$ is nonempty. Assume it is empty for some $T\in \mathcal{T}$, then by definition $|T|>1$. Since, $x\mathbf{R}^cy$ entails $x\not=y$, this means that there exists a finite sequence $(x_i)\in T^n$ such that $x_n\mathbf{R}^c x_{n-1}\mathbf{R}^c\dots\mathbf{R}^c x_1 \mathbf{R}^cx_n$. Hence, $x_n\prec x_{n-1}\prec\dots\prec x_n$, which contradicts the transitivity of $\succ$ on $T$. Therefore, we can define $x^\star_T = \min  \big(\{x \in T \mid \nexists y, x\mathbf{R}^c y\},\succsim\big)$.

Note that for each $T\in \mathcal{T}$ such that $|T|>1$, $x_T^\star\succ x_T$. Indeed, suppose by contradiction that $x_T\succsim x_T^\star$. Then for no $y\in T$, $x_T\mathbf{R}^cy$. But since $|T|>1$, this would mean that there exists $z\in T$ such that $z\mathbf{R}^cx_T$. By Lemma \ref{Lemmarelou2}, this entails $z\rhd_T x^T$, which contradicts the fact that $x^T$ is $\unrhd_T$-maximum. 

We now define, for each $T\in \mathcal{T}$, the binary relation $\tilde{\unrhd}_T\subseteq T^2$ as follows:  $\tilde{\unrhd}_T$ is equal to $\unrhd_T$ whenever $\unrhd_T$ is asymmetric; and for any pair $x,y \in T$ such that $x\bowtie_T y$ and $x\succ y$, we set $y ~ \tilde{\unrhd}_T~ x$ whenever $x_T^\star\succsim x\succ  y\succsim x^T$ and $x ~ \tilde{\unrhd}_T ~ y$ otherwise. In other words, whenever $x\bowtie_T y$, $\tilde{\unrhd}_T$ is equal to $\succsim$, except if both $x_T^\star \succsim x \succsim x^T$ and $x_T^\star \succsim y \succsim x^T$

One can verify that $\tilde{\unrhd}_T$ is complete and antisymmetric. We now prove that $\tilde{\unrhd}_T$ is transitive. Let $x,y,z\in T $ be such that $x~\tilde{\rhd}_T~y~\tilde{\rhd}_T~z$ (the cases where $x=y$ or $y=z$ are trivial, hence we only cover the cases where $x\neq y \neq z$). If $\neg [y\bowtie_Tx]$ or $\neg [z\bowtie_Ty]$, then $ x~\tilde{\rhd}_T~z$ follows from the transitivity of $\rhd_T$. Assume now that $x\bowtie_T y \bowtie_Tz$. Note that by transitivity of $\unrhd_T$, then  $x\bowtie_Tz$. Let us consider the possible cases:
\begin{enumerate}[(i)]
    \item $x\prec y \prec z$. Then, $x~\tilde{\rhd}_T~y~\tilde{\rhd}_T~z$ imply that $x^T \precsim x,y,z \precsim x^\star_T$. Furthermore, the transitivity of $\succsim$ implies that $x\prec z$. Hence, $x~\tilde{\rhd} ~z$. 
    \item $x\succ y \succ z$. Then, the transitivity of $\succsim$ implies that $x \succ z$. Suppose by contradiction that $z \ \tilde{\unrhd}_T \ x$, this implies that $x^T \precsim x,z \precsim x^\star_T$. Given that $x\succ y \succ z$ and by transitivity, it implies that $x^T \precsim y \precsim x^\star_T$, from which we should conclude that $z~\tilde{\unrhd}_T ~ y$ and $y~\tilde{\unrhd}_T ~ x$, a contradiction. Hence, $x~\tilde{\rhd} ~z$. 
    \item $x\succ y \prec z$. Then, $y ~\tilde{\rhd}_T ~ z$ implies that $x^T \precsim y,z \precsim x^\star_T$. Thus, $x ~\tilde{\rhd}_T ~ y$ and the transitivity of $\succsim$ imply that $x \succ x^\star_T$, implying in turn that $x \succ z$. Hence, $x~\tilde{\rhd} ~z$. 
    \item $x\prec y \succ z$. Then $x ~\tilde{\rhd}_T ~ y$ implies that $x^T \precsim x,y \precsim x^\star_T$. Thus, $y ~\tilde{\rhd}_T ~ z$ and the transitivity of $\succsim$ imply that $x^T \succ z$, implying in turn that $x \succ z$. Hence, $x~\tilde{\rhd} ~z$. 
\end{enumerate}

We finally define the relation $\succsim_2$ on $X$ as follows: for any $x,y \in X$,
\begin{align}\label{KeyRelation2}
    x\succsim_2 y\iff \left\{\begin{array}{ll}
    x ~\tilde{\unrhd}_T ~ y    & \text{if } \mathcal{T}(x) = \mathcal{T}(y) = T \\
    x\succsim y    & \text{otherwise}
   \end{array} 
   \right.
\end{align}
As usual, $\succ_2$ is its asymmetric part. 

One can verify that $\succsim_2$ is complete and antisymmetric. We now prove that it is transitive. Let $x,y,z \in X$ be such that $x\succ_2 y\succ_2 z$ (the cases where $x=y$ or $y=z$ are trivial, hence we only cover the cases where $x\neq y \neq z$). The only cases that have not been covered yet are the cases where exactly two options out of $\{x,y,z\}$ are from the same element $T \in \mathcal{T}$. If $\mathcal{T}(x)=\mathcal{T}(y)\not=\mathcal{T}(z)$, then $y\succ z$ and $x\unrhd_{\mathcal{T}(x)} y$ which, by definition of $\unrhd_{\mathcal{T}(x)}$ yields $x\succ z$, i.e., $x\succ_2 z$. 
The same reasoning can be applied if  $\mathcal{T}(x)\not=\mathcal{T}(y)=\mathcal{T}(z)$. Now if $\mathcal{T}(y)\not=\mathcal{T}(x)=\mathcal{T}(z)$, then $x\succ y\succ z$. Hence, it cannot be that $z~\unrhd_{\mathcal{T}(x)}~x$ and  by completeness of $\unrhd_{\mathcal{T}(x)}$, we have $x\rhd_{\mathcal{T}(x)}z$. Hence, $x~\tilde{\rhd}_{\mathcal{T}(x)}~z$, i.e., $x\succ_2 z$.

The relation $\bigcup_{T\in\mathcal{T}}\succsim_{|T}$ constitutes a partial order. Hence, by Szpilrajn's Extension Theorem, there exists a linear order $\succsim_1$ that completes this relation. 

We finally prove that $\langle \mathcal{T}, \succsim_1, \succsim_2 \rangle$ is a RS-structure that rationalizes $c$. Since $\succsim_1$ is transitive, we can define, for any menu $A$,
\begin{align*}
    d(A) = \{x \mid x = \max (\mathcal{T}(x) \cap A, \succsim_1)\}. 
\end{align*} 
For any $x,y \in d(A)$, $\mathcal{T}(x )\neq \mathcal{T}(y)$, so $\succsim_2\cap d(A)^2 =  \succsim \cap \ d(A)^2$, and
\begin{align*}
    \max\big(d(A),\succsim_2\big) = \max\big(d(A),\succsim\big).
\end{align*}
The following Lemma thus completes the proof of Theorem \ref{thm1}. 
\begin{lemma}
    For all $A\subseteq X$, $\max\big(d(A),\succsim\big) = c(A)$
\end{lemma}

\begin{proof}
For any menu $A$, let $i(A)$ be the number of types $T\in\mathcal T$ such that $T\cap A\not=\emptyset.$ If $i(A)=1$, $d(A)$ is a singleton, and the conclusion follows from the transitivity of $\succsim$ over each $T$ and \textbf{Exp}. 

Assume now that $i(A) = 2$. Let $x, y\in A$ be such that $\mathcal{T}(x) \not = \mathcal{T}(y)$, $x = \max (\mathcal{T}(x) \cap A, \succsim)$, $y = \max (\mathcal{T}(y) \cap A, \succsim)$, and $y \succ x$. Assume by contradiction that $y \neq c(A)$. By definition of $y$, $y\succ z$ for any $z \in \mathcal{T}(y) \cap A$. Hence, there must exist $z \in \mathcal{T}(x)$ such that $z \succ y $ and $y \not= c\{x,y,z\}$, otherwise \textbf{Exp} would imply that $y = c(A)$. This implies that $x\succ z \succ y \succ x$. Since $y \not= c\{x,y,z\}$,  either $y\mathbf{R}^cz$ and $\mathcal{T}(y)=\mathcal{T}(z)$, or $x\mathbf{R}^c y$ and $\mathcal{T}(x)= \mathcal{T}(y)$, both of which contradict that $\mathcal{T}(x)=\mathcal{T}(z)\not= \mathcal{T}(y)$. Hence, we conclude that $y=c(A)$.

Finally, fix $3 \leq n \leq n^*+1$ and let $A$ be a menu such that $i(A)=n$. We denote $y=\max (d(A),\succsim)$. Given the preceding proof, for any $z \in A$, $y = c\bigg(\big(\mathcal{T}(y) \cup \mathcal{T}(z)\big) \cap A\bigg)$. This implies by \textbf{Exp} that $y=c(A)$.
\end{proof}
\end{proof} 

\subsection{Proof of Theorem \ref{THMSinglePeaked}}
\label{app:proof_SPR}

\begin{proof}[Necessity of the axioms]
    Assume that $c$ is a single-peaked RSC, and for any $T \in \mathcal{T}$, let $x_T^{\star} \in T$ be such that: $\succsim_2 = \succsim_1$ on $\left[x_T^{\star}, \max(T,\succsim_1)\right]^{\succsim_1}$; and $\succsim_2$ is single-peaked with respect to $\succsim_1$ on $\left[\min(T,\succsim_1), x_T^{\star}\right]^{\succsim_1}$. Consider $x,y,z,t\in T$ such that $ x\succ y \succ z$, $x\mathbf{R}^ct$ and $z\mathbf{R}^cy$. Proposition \ref{RSC_property} implies that $z\succ_2 y$ and $y\succ_1z$. Furthermore, $x\succ y$ implies that $x\succ_1 y$; hence, $x \succ_1 z$ by transitivity of $\succsim_1$. Since, $x\mathbf{R}^ct$, $x\in [\min(T,\succsim_1),x^{\star}_T]^{\succsim_1}$, and thus $y,z\in [\min(T,\succsim_1),x^{\star}_T]^{\succsim_1}$, again by transitivity of $\succsim_1$. Since $z\succ_2 y$, the single-peakedness property implies that $y\succ_2 x$. Therefore, we obtain as desired that for all $u\notin \mathcal{T}(x)$: $x\succ u\iff x\succ_2 u\implies y\succ_2 u\iff y\succ u$. 
\end{proof} 

\begin{proof}[Sufficiency of the axioms]
    We already know from Theorem \ref{thm1} that $c$ is an RSC. Let $\langle \mathcal{T}, \succsim_1, \succsim_2 \rangle$ be the rationalizing RS-structure constructed in the proof of Theorem \ref{thm1}. We also refer to each object defined in the proof of Theorem \ref{thm1} using the same notation. Remember that within a type $T \in \mathcal{T}$, $\succsim_1 =\succsim$ and $\succsim_2 = \tilde{\unrhd}_T$, hence we use those notations interchangeably.
    
    We first establish the following useful lemma. 
    \begin{lemma}\label{Lemmarelou1}
        For any $T\in \mathcal{T}$ such that $|T|>1$ and any $x\in [\min(T,\succsim_1),x^\star_T]^{\succsim_1}\setminus\{x^\star_T\}$, $x\mathbf{R}^cx_T^\star$. 
    \end{lemma}
    \begin{proof} 
        Let $x \in [\min(T,\succsim_1),x^\star_T]^{\succsim_1}\setminus\{x^\star_T\}$. Note first that by definition of $\succsim_1$ and $x^\star_T$, for any such $x$, there must exist $x_1 \in T$ such that $x\mathbf{R}^c x_1$. If $x_1\succsim_1x^\star_T$, then $x_1\succsim_2x_T^\star$, and therefore, Proposition \ref{RSC_property} implies the existence of some $t\notin T$ such that $x \succ_2 t \succ_2 x_1 \succsim_2 x^\star_T$, which would imply in turn that $x\mathbf{R}^cx_T^\star$. Conversely, suppose that $x_T^\star \succ_1 x_1$. Then, it implies that there exists $x_2\in T$ such that $x_1\mathbf{R}^cx_2$, otherwise $x_T^\star$ is not a $\succ_1$-minimizer over $\{x\in T:\nexists y\in T, x\mathbf{R}^cy\}$. Similarly, if $x_2 \succsim_1  x_T^\star$, then we would obtain that $x_1 \mathbf{R}^c x^\star_T$, thus concluding that $x \mathbf{R}^c x^\star_T$ from the transitivity of $\mathbf{R}^c$ (which simply follows from the transitivity of $\rhd_T$, $\succ$ on $T$, and Lemma \ref{Lemmarelou2}). Conversely, if $x^\star_T \succ_1 x_2$, then there exists $x_3$ such that $x_2 \mathbf{R}^c x_3$. By iterating the same argument we obtain an infinite sequence $(x_n)_n$ of different options in $X$ such that $x_{n-1} \mathbf{R}^c x_n$, contradicting the finiteness of $X$. 
    \end{proof}
    We show that for each $T\in \mathcal{T}$, $\succsim_2=\succsim_1$ on $[x^\star_T,\max(T,\succsim)]^{\succsim_1}$. If $|T|=1$, there is nothing to prove. If $|T|>1$, then we know from Lemma \ref{Lemmarelou1} that $x^T \mathbf{R}^c x^\star_T$, and by definition there is no $y$ such that $x^\star_T\mathbf{R}^c y$. \textbf{SPR} thus implies that for any $x \in T$ such that $x\succ x^\star_T$, there is no $y$ such that $x\mathbf{R}^c y$. Indeed, if this were the case, \textbf{SPR} would imply that $x^\star_T \unrhd_T x$, and thus $x^\star_T\mathbf{R}^c y$, by Lemma \ref{Lemmarelou2}, a contradiction. Given that for any $x,y \in T$ such that $y\succ x \succsim x^\star_T$, then $\neg[x\mathbf{R}^cy]$, this means that $y\unrhd_T x$, and thus by definition of $\tilde{\unrhd}_T$, $y~\tilde{\rhd}_T ~x$, i.e., $y\succ_2 x$. 
    
    Now, let $x,y,z \in [\min(T,\succsim_1),x^\star_T]^{\succsim_1}$ be such that $x\succ_1 y\succ_1 z$, and assume that $z \succ_2 y$, that is, $z~\tilde{\rhd}_T ~y$. We want to show that $y\succ_2 x$, that is, and $y~\tilde{\rhd}_T~ x$. If $x=x_T^\star$, Lemma \ref{Lemmarelou1} entails that $y\mathbf{R}^cx$, and thus $y~\tilde{\rhd}_T~ x$, as Lemma \ref{Lemmarelou2} implies that $y\rhd_T x$. 
    
    Suppose now that $x \prec_1 x_T^\star$, which implies that $x\mathbf{R}^ct$ for some $t\in T$. First, $z\succ_2 y$ implies that $z\unrhd_T y$. We first prove that $y\succsim x^T$. Suppose by contradiction that $y\prec x^T$. In this case, from the definition of $\tilde{\unrhd}_T$, $z~\tilde{\rhd}_T ~y$ implies that $z\rhd_T y$, and thus, by Lemma \ref{Lemmarelou2}, $z\mathbf{R}^c y$. \textbf{SPR} implies that $y \unrhd_T x^T$, which contradicts the definition of $x^T$. Therefore, $y \succsim x^T$.

    If $y\bowtie_T x^T$, then by definition of $x^T$ and transitivity of $\unrhd_T$, $y\unrhd_T x$. If $x^T \rhd_T y$, then it means that $x^T \mathbf{R}^c y$ (Lemma \ref{Lemmarelou2}). In this case, \textbf{SPR} implies that $y\unrhd_T x$. Given that $x^T \precsim y\prec x \prec x^\star_T$, the definition of $\tilde{\unrhd}_T$ implies that $y~\tilde{\rhd}_T x$, i.e., $y\succ_2 x$, which concludes the proof.
\end{proof}


\section{Proof of Theorem \ref{menu_pref}}\label{proof_menu}

\begin{proof}
The proof of the necessity of the axioms is left to the readers. We only prove their sufficiency. Define $F =\bigcup_{T\in \mathcal{T}} F^T$.

(a) We first show that for any $A, B$ such that $A \subseteq T$ and $B \subseteq T'$ for some $T,T' \in \mathcal{T}$, $A \succ B \iff A \cap F \neq \emptyset = B \cap F$. Note that $A \cap F \neq \emptyset = B \cap F$ implies that $A$ is richer than $B$ and by R-Dominance (RD) $A\succ B$. Now we assume by contradiction that $A \cap F \neq \emptyset = B \cap F$ does not hold and show that then $B\succsim A$.  Denote $A'= A \setminus F = \{a_1,\dots, a_n\}$ and $B'= B \setminus F = \{b_1, \dots, b_l\}$ and suppose that both are non-empty. By RD, $\{a_1\} \sim A'$, because both are richer than each other. Similarly $\{b_1\} \sim B'$. Furthermore, RD \textit{(ii)} implies that $\{a_1\}\sim \{b_1\}$; hence, by transitivity, $A' \sim B'$. Denote $A''=A \setminus A'$ and $B''=B \setminus B'$. If $A''=B''=\emptyset$, we conclude from the previous argument that $A\sim B$. Suppose finally that $B''\neq \emptyset$, and possibly $B'=\emptyset$. Then, $A''=\emptyset $ implies by RD that $B\sim B''\succ A$. Hence transitivity implies $B\succ A$. By RD, $A\sim A''$ because they are both richer than each other, and similarly $B\sim B''$. By a similar reasoning as for $A'$ and $B'$, one can show that $A''\sim B''$, and thus $A\sim B$ results from transitivity. 

(b) We next show that for any $A, B$, if $\#\Phi(A) = \#\Phi(B)$, then $A\sim B$. For any menu $D \in \mathcal{X}$, we define $\Phi(D) = \{F^T\cap D : T \in \mathcal{T} \text{ and } F^T\cap D \neq \emptyset \}$. Denote $\Phi(A) =\{A_1,\dots, A_n\}$ and $\Phi(B)=\{B_1,\dots, B_n\}$. By (a), we know that for any $i$, $A_i \sim B_i$. Noting that $A_1 \cap A_2 = B_1 \cap B_2 = \emptyset$, and neither $A_1$ is richer than $A_2$ nor $B_1$ is richer than $B_2$, by applying twice R-Composition, we get that $A_1 \cup A_2 \sim B_1 \cup B_2$. By reiterating the same argument, we obtain that $\bigcup_i A_i \sim \bigcup_i B_i$. Finally, note that $A$ is richer than $\bigcup_i A_i$ and conversely $\bigcup_i A_i$ is richer than $A$, hence, by RD, $A\sim \bigcup_i A_i$; similarly $B \sim \bigcup_i B_i$. Therefore, by transitivity, we obtain that $A \sim B$. 

(c) We finally prove that for any $A, B$, if $\#\Phi(A) > \#\Phi(B)$, then $A\succ B$. Denote $\Phi(A) =\{A_1,\dots, A_n\}$ and $\Phi(B)=\{B_1,\dots, B_k\}$, with $k<n$. By (b) $\bigcup_{i=1}^k A_i \sim B$. Furthermore, by RD, $\bigcup_{i=1}^n A_i \succ \bigcup_{i=1}^k A_i$. A similar argument as before shows that $A \sim \bigcup_{i=1}^n A_i$ and $B \sim \bigcup_{i=1}^k B_i$. Hence, by transitivity, $A \succ B$.
\end{proof}


\section{Proofs of Section \ref{applications}}\label{proof_appli}

\begin{proof}[\textbf{Proof of Proposition} \ref{conspiracy}]
\textit{(i)} In the menu $N$, the DM compares $\sigma^L$ and $\sigma^{RR}$ in the following way: 
\begin{align*}
    \sigma^L\prec_2\sigma^{RR} & \iff (1-p)+p\lambda -p(1-\lambda) \leq p + (1-p)\delta \\
            & \iff p \geq \frac{1-\delta}{3-2\lambda-\delta}=\frac{1/2}{5/2-2\lambda}
\end{align*}
We define $p^\star:=\frac{1/2}{5/2-2\lambda}$ and verify that $p^\star<1/2$:
\begin{align*}
    p^\star<1/2 & \iff \lambda < \frac{3}{4}
\end{align*}
which is true by assumption.

\textit{(ii).} We first compute the value $q^\star$ of the posterior such that for any $q \geq q^\star$, action $r$ is preferred. $q^\star$ solves $(1-q) - q = q$, hence $q^\star = 1/3$. 

Then we simply compare the posterior obtained after the realisation of a signal $s^r$ from the news source $\sigma^{RR}$ with $1/3$. The posterior is, $\frac{p}{p + (1-p)1/2}$, which is greater or equal than $p$. We are in the case where $p \geq p^\star$, hence it is sufficient to show that $p^\star \geq 1/3$: $p^\star \geq \frac{1}{3} \iff \lambda \geq \delta$
which is true by assumption.
\end{proof}

\begin{proof}[\textbf{Proof of Lemma} \ref{Connec}]
The maximand of the program (\ref{program35}) is concave and the set $K_g$ is compact and bounded, hence, $c$ is well-defined (Weierstrass theorem). Furthermore, from (\ref{FOC}) we see it has a unique solution, thus $c$ is a choice function.

Now let us build the RS-structure $\langle \mathcal{T}, \succsim_1, \succsim_2 \rangle$ that represents $c$. Given (\ref{FOC}), $d^\star$ strictly increases with $g$ if and only if $g \in [\hat{g},\Bar{g}]$. Let $g(t,d)$ be the $g$ such that $t+gd^\beta=1.$

Let us introduce the three following sets:
\begin{align*}
    & D^\uparrow=\bigcup_{t\in[0,1]\atop g> \hat{g}}\{d\in [0,1]:(t,d)=c(K_g)\},\\
    & \forall d\in D^\uparrow,\mathcal{T}(d)=\bigcup_{ t\in [0,1]}\{(t,d)\}, \\
    & T_0^c=\bigcup_{d\notin D^\uparrow \atop t\in [0,1]}\{(t,d)\}.
\end{align*}
From these sets, we can define the set of types and the thresholds: 
\begin{align*}
    &\mathcal{T}=\{T_0^c\}\cup\bigcup_{d\in D^\uparrow}\{\mathcal{T}(d)\}, \\
    \text{ and for any } & T\neq T_0^c, \ T= \mathcal{T}(d), \quad  \{x^\star_T\}=\{(t,d)\in T : g(t,d)= \hat{g}\}.
\end{align*}
For each $(t,d)$ we posit $u(t,d)=t+P(d)\hat{V}$, $v(t,d)=u(t,d)$ if $(t,d)\in T_0^c$ or $u(t,d)\geq u(x^{\mathcal{T}(d)})$, and $v(t,d)=t+P(d)V\big(g(t,d)\big)$ otherwise. Given the uniqueness of $g(t,d)$ for each $(t,d)$, $v$ is well-defined. Defining $\succ_1$ and $\succ_2$ as the preference representing $u$ and $v$. It can easily be checked that $\langle \mathcal{T}, \succsim_1, \succsim_2 \rangle$ is an RS-structure. Consider the RSC $c'$ defined on the collection of menus $\{K_g\}_{g\geq 1}$ and represented by $\langle \mathcal{T}, \succsim_1, \succsim_2 \rangle$. 

Note that for any $(t,d) \in [0,1]^2$ and $g\geq 1$, $(t,d)\in K_g \iff g(t,d) \geq g$. Therefore for any $T \neq T_0$, $x^\star_T \in K_g \iff g(x^\star_T) \geq g$. This means that for any $g\geq 1$ and $d \in D^\uparrow$, there exists $t$ such that $(t,d)\in K_g$ and $u(t,d) \geq u(x^{\mathcal{T}(d)})$ if and only if $g\leq \hat{g}$. Given the definitions of $u$, and $v$, this implies for any $g\geq 1$, $c(K_g)=c'(K_g)$. 
\end{proof}

\begin{proof}[\textbf{Proof of Proposition} \ref{dydy2}]
This is a straightforward consequence of (\ref{dydy}).
\end{proof}

\section{Relation with TSM}
\label{app:TSM}

We show that the model TSM \citep{horan_2016} violates our axioms \textbf{NRC} and \textbf{SPR}.
 
\begin{definition}\label{tsm_def}
    A TSM is a pair $(P_1, P_2)$ of transitive and asymmetric binary relations on $X$ that defines a choice function $c$ as follows: for any menu $A$,
    $$ c(A) = \max (\max(A;P_1), P_2); $$
    where $\max(A;P) = \{x \in A: \nexists y \in A \emph{ s.t. } yPx\}$.
\end{definition}
 
To see how that TSM violates \textbf{NRC}, consider the following example:

\begin{center}
    \begin{tabular}{c|c|c}
	$X$ & $P_1$ & $P_2$ \\ \hline
	$\{x,y,z,t,u\}$ & $zP_1xP_1t$, $yP_1t$ & $tP_2uP_2xP_2yP_2 z$
    \end{tabular}
\end{center}
Then, $t=c\{u,t\}$ and $u=c\{x,u,t\}=c\{y,u,t\}=c\{z,u,t\}$. This implies $t\mathbf{R}^cx$, $t\mathbf{R}^cy$, $t\mathbf{R}^cz$, i.e. $x\approx y \approx z$. However, $x=c\{x,y\}$, $y=c\{y,z\}$, and $z=c\{x,z\}$, violating \textbf{NRC}. 

To see that TSM violates \textbf{SPR}, consider the following example:
	
\begin{center}
    \begin{tabular}{c|c|c}
	$X$ & $P_1$ & $P_2$ \\ \hline
	$\{x,y,z,a,t\}$ & $tP_1 z P_1 y P_1 x$ & $zP_2 x P_2 a P_2 t P_2 y$
    \end{tabular}
\end{center}	
First, note that $z=c\{y,z\}$ and $y=c\{x,y\}$. Furthermore, one can check that $x\mathbf{R}^c t$, $z \mathbf{R}^c t$, and $x \mathbf{R}^c y$. However, one can also check that $a\not\approx z$, $z=c\{a,z\}$ while $a=c\{a,y\}$, which contradicts \textbf{SPR}.

\end{appendices}

\bibliography{Bibliography}
\bibliographystyle{chicago}

\end{document}